\documentclass[aps,pre,showpacs, amsmath,  amssymb,reprint,groupedaddress,nofootinbib,10pt]{revtex4-1}
\usepackage{amsthm}
\usepackage{color}
\usepackage{graphicx}
\usepackage{verbatim}

\newcommand{\theTitle}{Transfer entropy in continuous time, with applications to jump and neural spiking processes}

\newcommand{\theAuthors}{Richard E. Spinney, Mikhail Prokopenko and Joseph T. Lizier}

\newtheorem{prop}{Proposition}
\newtheorem{definition}{Definition}
\newtheorem{corol}{Corollary}
\newtheorem{remark}{Remark}

\newcommand{\secRef}[1]{Sect.~\ref{sec:#1}}
\newcommand{\eq}[1]{eq.~(\ref{eq:#1})}
\newcommand{\Eq}[1]{Eq.~(\ref{eq:#1})}

\mathchardef\mhyphen="2D
\begin{document}
\title{\theTitle}
\author{\theAuthors}

\affiliation{Complex Systems Research Group and Centre for Complex Systems, Faculty of Engineering \& IT, The University of Sydney, NSW 2006, Australia}
\date{\today}

\begin{abstract}
Transfer entropy has been used to quantify the directed flow of information between source and target variables in many complex systems.
While transfer entropy was originally formulated in discrete time, in this paper we provide a framework for considering transfer entropy in continuous time systems, based on Radon-Nikodym derivatives between measures of complete path realizations. 
To describe the information dynamics of individual path realizations, we introduce the pathwise transfer entropy, the expectation of which is the transfer entropy accumulated over a finite time interval. We demonstrate that this formalism permits an instantaneous transfer entropy rate. These properties are analogous to the behavior of physical quantities defined along paths such as work and heat.
We use this approach to produce an explicit form for the transfer entropy for pure jump processes, and highlight the simplified form in the specific case of point processes (frequently used in neuroscience to model neural spike trains).
Finally, we present two synthetic spiking neuron model examples to exhibit the pertinent features of our formalism, namely, that the information flow for point processes consists of discontinuous jump contributions (at spikes in the target) interrupting a continuously varying contribution (relating to waiting times between target spikes). Numerical schemes based on our formalism promise significant benefits over existing strategies based on discrete time formalisms. 
\end{abstract}
\maketitle
\section{Introduction}
The nature of directed information flow between entities in distributed complex systems is of wide interest across neuroscience, economics, systems biology, multi-agent systems, etc.
To quantify the directed information flow between two variables based on samples of time-series of their activity, the \emph{transfer entropy} \cite{schr00} has become the standard approach \cite{liz10a,liz14c}.
Transfer entropy \cite{schr00,palus01} from $Y$ to $X$ for a pair of coincident time ordered sequences $({{x}}_0^n,{{y}}_0^n)$ where ${{x}}_0^n=({x}_0,{x}_1\ldots {x}_n)$ such that the subscript $n$ is a (discrete) time index, is given by
\begin{align}
T^{(k,l)}_{{y}\to{x}}(n)&=\left\langle \ln{\frac{P({x}_{n}|{{x}}^{n-1}_{n-k},{{y}}^{n-1}_{n-l})}{P({x}_{n}|{{x}}^{n-1}_{n-k})}}\right\rangle
\label{eq:def} \\
&= \left\langle t^{(k,l)}_{{y}\to{x}}(x_n,{{x}}^{n-1}_{n-k}, {{y}}^{n-1}_{n-l}) \right\rangle
\label{eq:defLocal},
\end{align}
measured here and throughout in nats, where $\{k>0\}\in\mathbb{N}, \{l>0\}\in\mathbb{N}$. Here, $\langle\ldots\rangle$ indicates an ensemble average and we use $P$ to indicate a probability distribution for an implied discrete random variable $x$.
Note that $t^{(k,l)}_{{y}\to{x}}$, the local transfer entropy \cite{liz08a}, represents the log ratio for a given sample $\{ {{x}}^{n}_{n-k},{{y}}^{n-1}_{n-l} \}$ at time $n$.
Indeed, one may interpret the local quantity as the difference in ``surprise'' \cite{cover91} of the outcome $x_n$ between scenarios where the history of $x$ is known and where the history of $y$ is known in addition to that of $x$. This difference thus characterizes how helpful the history of $y$ was in predicting $x_n$ over and above the history of $x$.
Transfer entropy may be suitably extended for continuous random variables $x$ and $y$ (in discrete time) by replacing discrete probabilities $P$ with probability density functions $p$, with appropriate weighted integrals over $p({x}_{n}|{{x}}^{n-1}_{n-k},{{y}}^{n-1}_{n-l})$ for the expectation value \cite{kais02}, based on underlying differential entropies \cite{cover91}.

The transfer entropy is a measure of predictive information transfer, not of causal effect \cite{liz10a}. It is particularly useful in describing distributed information processing (where raw causality is not), such as identifying emergent dynamic structures (i.e., gliders) in cellular automata \cite{liz08a}, cascading information waves in swarms \cite{wang12a}, and information carrying signal properties in biochemical pathways \cite{pahle08}.

Indeed, transfer entropy has proven particularly popular in computational neuroscience for characterizing neural information flows, with applications such as inferring effective neural information networks underpinning cognitive tasks and their variation \cite{hon07,lind11a,woll14a,stett12a,faes15a}, across data modalities including magnetoencephalography (MEG) \cite{vic11a,wib11a}, electroencephalography (EEG) \cite{stra12b,mar12c,stra14a}, and functional magnetic resonance imaging (fMRI) \cite{liz11a,diez15a}.
Applications to spike train data have been less abundant, however.
This is because in considering neural spiking data, and many other processes ubiquitous in fields ranging from physics to economics and beyond, we do not have a discretized time basis, but instead have events which occur at an arbitrary resolution in continuous time.
How should one rigorously compute the transfer entropy for such data sets?
Previous approaches have attempted to apply the discrete time formalism to such systems in a number of ways, for example, in examining the information between most recent events in an economic setting \cite{harre15a}, or in discretizing time (i.e., time binning) for spiking neural processes \cite{gour07,ito11a,tim14a,timm16a,thiv14a}. 
Such approaches necessarily recast the dynamics in order to make empirical approximations, which may ignore key mechanisms relevant to the source-target relationship.
In particular, discretizations in time cannot detect interactions (including feedback) below the resolution of the discretization.
Choosing a fine discretization (e.g. to ensure only one event occurs in any bin, requiring, for example in neural spike processes, bin sizes of ms order or less) to counter this, however, leads to the temporal history either being seriously undersampled or simply ignored.\footnote{Employing a fine discretization will lead to values for $k$ in ${{x}}^{n-1}_{n-k}$ becoming impractically large since, for example, the temporal structure in spike trains is often tens or even several hundred ms long \cite{ald91,bargad01}, or indeed scale-free for critical dynamics \cite{beggs03,lev07a,prie13a,prie14a,rubi11a}.}
While it may be possible to optimize such trade-offs for a time discretization (e.g. \cite{bor16a}), one cannot simultaneously avoid all of these issues in general.

Instead, we argue that optimal treatment of information flow in these processes (such as neural spike trains) first requires a distinct theoretical understanding of the nature of transfer entropy in continuous time. Such an approach requires rethinking the idea that transfer entropy is a quantity that is defined at an instant in time for which there are local versions, but rather considers transfer entropy as a quantity that is accumulated over a finite time interval with an associated instantaneous transfer entropy rate. The transfer entropy accumulated over a time interval is the average of an individual fluctuating quantity along a single path realization which we call the pathwise transfer entropy. The pathwise transfer entropy for a given realization is not guaranteed to be smooth meaning that even where an (average) transfer entropy rate exists, the notion of a local transfer entropy rate, at a given instant of an individual realization, may not generally be well defined.

We begin by presenting how the transfer entropy should be reconsidered in continuous time (\secRef{contTime}) wherein we present our central quantities, the transfer entropy rate and pathwise transfer entropy, which, in order to be expressed generally require a measure-theoretic formulation.
We then apply this formalism, offering analytic forms for our central quantities, for jump processes (those which exhibit jumps between states at continuous time points) in \secRef{jump}, and specialize this solution for neural spike trains, or more broadly point processes, in \secRef{spikes}. 
Next, we apply our solution for transfer entropy for spike trains to a number of scenarios in \secRef{examples}, in order to highlight the properties of the approach and how results should be interpreted.
Our results imply a simple empirical form for the transfer entropy rate for spike trains, summing -- at each target spike only -- the log ratio of history-dependent spike rates, with and without knowledge of the source.
We expect these results to have significant influence on the measurement of information transfer in data sets from point processes, for example, broadening the already wide application of transfer entropy in computational neuroscience to spike train data sets.

\section{Transfer Entropy in continuous time}
\label{sec:contTime}
\subsection{Measure-theoretic transfer entropy}
In this section, we establish a generalized form for the transfer entropy in terms of relationships between probability measures on arbitrary stochastic processes.
To do so, we utilize measure-theoretic approaches, an oft-quoted rationale for which is to unify the \emph{ad hoc} methods which exist for discrete and continuous random variables and, under one framework, allow for the discussion of random variables for which probability mass functions or densities cannot be readily formulated. For instance, these could be combinations of discrete and continuous random variables or more sophisticated quantities such as random fields.
While only a generalization in discrete time, this will be essential when we come to consider continuous time, where the complete behaviour of some process evolving in time is described by an uncountably infinite number of points, and so any formalism must be able to manage quantities which capture the whole process such as random functions.

Our first observation is that we can generalize \eq{def} by recognizing it as the expectation of the logarithm of the Radon-Nikodym derivative of a given conditional probability measure with respect to a distinct, but equivalent,\footnote{Equivalent measures are those which are absolutely continuous with respect to the other such that each agree on which sets of events have probability zero.}, conditional probability measure (as observed in \cite{zhu15a}). We point out that the Radon-Nikodym derivative serves as the density of a measure with respect to another and can function as a generalized Jacobian, changing between those measures under an integral, analogously to a normal derivative. Heuristically, therefore, one may consider it to be the ratio of the probabilities assigned to a set in the relevant limit of that set size. For instance, in discrete time processes that concern finite state spaces it can be considered as the ratio of two different probabilities of a given event and for continuous state spaces it is the ratio of probability density functions.
In discrete time, this leads us to the following definition:
\begin{minipage}{\linewidth}
\vspace{1em}
\noindent\rule[0.5ex]{\linewidth}{0.5pt}
\vspace{-1em}
\begin{definition}
Given two stochastic processes $\{x_t\}_{t\in \mathbb{T}}$ and $\{y_t\}_{t\in \mathbb{T}}$ adapted to the underlying filtered probability space $(\Omega,\mathcal{F},\{\mathcal{F}_t\}_{t\in \mathbb{T}},P)$, indexed by the set of consecutive integers $\mathbb{T}\subseteq \mathbb{Z}$, the transfer entropy is the expectation of the logarithm of the Radon-Nikodym derivative between two equivalent measures on the random variable $x_n$ taking values in a measurable state space $(\Sigma_x,\mathcal{X})$, which are regular conditional probabilities given two related conditions:
\begin{align}
T_{y\to x}^{(k,l)}\big|_{n-1}^n&=\mathbb{E}_{P}\left[\ln{\frac{d\mathbb{P}_n(x_n|x^{n-1}_{n-k},y^{n-1}_{n-l})}{d\mathbb{P}_n(x_n|x^{n-1}_{n-k})}}\right]\nonumber\\
&=\int_{\Omega}\ln{\frac{d\mathbb{P}_n(x_n|x^{n-1}_{n-k},y^{n-1}_{n-l})}{d\mathbb{P}_n(x_n|x^{n-1}_{n-k})}(\omega)}dP(\omega).
\label{measure}
\end{align}

The extended notion of a local transfer entropy is analogously defined as
\begin{align}
\mathcal{T}_{y\to x}^{(k,l)}(x_n,x_{n-k}^{n-1},y_{n-l}^{n-1})&=\ln{\frac{d\mathbb{P}_n(x_n|x^{n-1}_{n-k},y^{n-1}_{n-l})}{d\mathbb{P}_n(x_n|x^{n-1}_{n-k})}}.
\label{localmeasure}
\end{align}
\label{radonDef}
\end{definition}
\noindent\rule[0.5ex]{\linewidth}{0.5pt}
\end{minipage}
With this notation for the transfer entropy, in contrast to Eq.~(\ref{eq:defLocal}), we introduce and emphasize the concept that this is the transfer entropy associated with, or accumulated over, the interval $n-1$ to $n$ indicated by the $\big|_{n-1}^{n}$ notation (see further discussion in \secRef{implicationsForEmpirical}). 
We may generalize this over longer intervals, in this instance $n-1$ to $n+m$, by writing:
\begin{align}
\mathcal{T}_{y\to x}^{(k,l)}&(x_{n}^{n+m},x_{n-k}^{n-1},y_{n-l}^{n-1+m}) \nonumber\\
= &\sum_{i=0}^{m} \mathcal{T}_{y\to x}^{(k,l)}(x_{n+i},x_{n-k+i}^{n-1+i},y_{n-l+i}^{n-1+i}),
	\label{localPathDiscrete}\\
	T_{y\to x}^{(k,l)}\big|_{n-1}^{n+m}&=\mathbb{E}_{P}\left[\mathcal{T}_{y\to x}^{(k,l)}(x_{n}^{n+m},x_{n-k}^{n-1},y_{n-l}^{n-1+m})\right].
\end{align}
For a stationary process this last line would be equal to $(m+1)T_{y\to x}^{(k,l)}\big|^{n}_{n-1}$. Explicitly, for Eq.~(\ref{measure}), in the special case that $x_n$ is a single discrete variable the integral w.r.t. the measure reduces to $(k+l+1)$ summations and the $d\mathbb{P}$ can be directly considered as probabilities [c.f. \eq{def}]. Similarly, in the case that $x_n$ is continuous the integral w.r.t. the measure reduces to $(k+l+1)$ integrals over a probability density w.r.t. $x^{n}_{n-k}$ and $y^{n-1}_{n-l}$ and the contents of the logarithm can, in entirety, be considered as the ratio between two probability densities.

In formalizing Definition \ref{radonDef} and other quantities, we shall make use of the following.
We consider $x$ and $y$, taking values in the measurable state spaces $(\Sigma_x,\mathcal{X})$ and $(\Sigma_y,\mathcal{Y})$ to be stochastic processes $\{x_t\}_{t\in \mathbb{T}}$ and $\{y_t\}_{t\in \mathbb{T}}$ adapted to the filtered probability space $(\Omega,\mathcal{F},\{\mathcal{F}_t\}_{t\in \mathbb{T}},P)$ with samples $\omega\in\Omega$ such that $x:\mathbb{T}\times\Omega\to\Sigma_x$. We assert the existence of the suitable measurable space $(\Omega^{\mathbb{T}}_x,\mathcal{F}^{\mathbb{T}}_x)$, where $\Omega^{\mathbb{T}}_x\subseteq(\Sigma_x)^{\mathbb{T}}$ such that samples are random functions, or \emph{paths}, $x_\mathbb{T}\equiv x(\cdot,\omega)$, i.e. $x_\mathbb{T}:\Omega\to\Omega_x^{\mathbb{T}}$. Similarly $(\Omega^{\mathbb{T}}_y,\mathcal{F}^{\mathbb{T}}_y)$ is the suitable measurable space for samples $y_\mathbb{T}\equiv y(\cdot,\omega)$. We equip these path spaces with a family of probability measures, denoted $\mathbb{P}_{X\cdot}^{\mathbb{T},(\cdot)}$, $\mathbb{P}_{Y\cdot}^{\mathbb{T},(\cdot)}$ (which we call natural measures), derived from the canonical (pushforward) measures, or laws, $\mathbb{P}_X^\mathbb{T}$ and $\mathbb{P}_Y^\mathbb{T}$ induced on $x_\mathbb{T}$ and $y_\mathbb{T}$. These canonical measures are the marginal measures of the probability space $(\Omega^{\mathbb{T}}_{xy},\mathcal{F}^{\mathbb{T}}_{xy},\mathbb{P}^{\mathbb{T}}_{XY})\equiv(\Omega^{\mathbb{T}}_x\times\Omega^{\mathbb{T}}_y,\mathcal{F}^{\mathbb{T}}_x\otimes\mathcal{F}^{\mathbb{T}}_y,\mathbb{P}^{\mathbb{T}}_{XY})$ induced on $z(\cdot,\omega)\equiv\{x(\cdot,\omega),y(\cdot,\omega)\}$. To recover and generalize the original definition of the transfer entropy in discrete time where $\mathbb{T}\subseteq\mathbb{Z}$ we consider the probability space $(\Sigma_x,\mathcal{X},\mathbb{P}_n)$ induced on the single random variable $x_n=x(n,\omega)$ [such that we also recognize $x^{n-1}_{n-k}=x^{n-1}_{n-k}(\omega)$]. 
By insisting that $\mathbb{T}\supseteq \{n-\max{(k,l)},\ldots,n\}$ we create the measures in Eq.~(\ref{measure}) in the manner of regular conditional probabilities \cite{billingsley_probability_1995}.\footnote{We understand $\mathbb{P}_n(x_n\in\mathcal{A}|x^{n-1}_{n-k})=P(x_n^{-1}(\mathcal{A})|x^{n-1}_{n-k})=P(\{\omega:x_n\in\mathcal{A}\}\in\mathcal{F}|x^{n-1}_{n-k})=\mathbb{E}_{P}[1_{\mathcal{A}}(x_n)|\sigma(x^{n-1}_{n-k})\subseteq\mathcal{F}_{n-1}]$ where $1_{\mathcal{A}}$ is the indicator function on $\mathcal{A}$ such that it satisfies $\mathbb{E}_{P}[1_{\mathcal{B}}(x^{n-1}_{n-k})\mathbb{P}_n(x_n\in\mathcal{A}|x^{n-1}_{n-k})]=\mathbb{P}(x_n\in \mathcal{A}\cap x^{n-1}_{n-k}\in \mathcal{B}),\quad\forall \;\mathcal{B}\in\sigma(x^{n-1}_{n-k})$ where $\sigma(\mathcal{G})$ denotes the sub-$\sigma$-algebra of $\mathcal{F}$ generated by $\mathcal{G}$.}
Given these measures we define the transfer entropy (in discrete time) as the expectation in Eq.~(\ref{measure}).
\subsection{Continuous time formalism}
To define such a quantity in continuous time we recognize that \eq{def} represents a rate of a transfer of information per discretized time step \cite{amb09a}. Consequently, without such a fundamental temporal discretization we must initially define a \emph{transfer entropy rate} in Proposition \ref{propcont} (see also \cite{hor14a,boss16a,barn16a,spin16a}). We emphasize that this naturally leads to integrated quantities, in the form of functionals of realized paths, which we introduce subsequently (Proposition \ref{pathdef}).
\begin{minipage}{\linewidth}
\vspace{1em}
\noindent\rule[0.5ex]{\linewidth}{0.5pt}
\vspace{-2em}
\begin{prop}
In continuous time such that we have stochastic processes $\{x_t\}_{t\in \mathbb{T}}$ and $\{y_t\}_{t\in \mathbb{T}}$, indexed by the connected subset $\mathbb{T}\subseteq \mathbb{R}$, we must consider the \emph{transfer entropy rate} which, analogously to Eq~(\ref{measure}), is given by
\begin{align}
\dot{T}^{(s,r)}_{{y}\to{x}}&(t)\nonumber\\
=&\lim_{dt\to 0}\frac{1}{dt}\mathbb{E}_{P}\left[\ln{\frac{d\mathbb{P}_{t+dt}[{x}_{t+dt}|{{x}}^{t}_{t-s},{{y}}^{t}_{t-r}]}{d\mathbb{P}_{t+dt}[{x}_{t+dt}|{{x}}^{t}_{t-s}]}}\right]\nonumber\\
=&\lim_{dt\to 0}\frac{1}{dt}\int_{\Omega}\ln{\frac{d\mathbb{P}_{t+dt}[{x}_{t+dt}|{{x}}^{t}_{t-s},{{y}}^{t}_{t-r}]}{d\mathbb{P}_{t+dt}[{x}_{t+dt}|{{x}}^{t}_{t-s}]}}(\omega)dP(\omega).
\label{eq:eq2}
\end{align}
\label{propcont}
\end{prop}
\vspace{-2em}
\noindent\rule[0.5ex]{\linewidth}{0.5pt}
\end{minipage}
The above uses notation convention $[\ldots]$ to indicate that arguments include path functions which we write using the notation $x_{t_0}^{t}=\{x(t',\omega): t_0 \leq t'< t\}$ and $x_t=x(t,\omega)$, where $t'\in \mathbb{T}\subseteq\mathbb{R}$. 
The conditional measures are constructed in the manner of regular conditional probabilities analogously to the discrete time case, but now conditional on previous path functions.
Expanding on \eq{eq2} in Proposition \ref{propcont}, we require $\mathbb{T}\supseteq[t-\max(s,r),t+dt]$ and introduce the variables $\{s>0\}\in \mathbb{R}, \{r>0\}\in \mathbb{R}$  which play the role of $k$ and $l$ in discrete time tuning how much previous history to use in the calculation. When they are omitted it is to be understood that it indicates the limit $s\to\infty$ and $r\to\infty$. We point out that Markovian dynamics are captured by the \emph{limit} $s\searrow 0, r\searrow 0$ (i.e. not $s=0$, $r=0$). We emphasize, in these forms, Eqs.~(\ref{measure}) and (\ref{eq:eq2}) allow for a very general application of transfer entropy since $x$ can represent any quantity which can be assigned a probability measure that evolves in time with the distinction simply being whether that evolution occurs in discrete or continuous time. \\
\\
We next introduce integrated versions of the transfer entropy which characterize the information transfer over finite time intervals through the use of probability measures on realizations of the stochastic processes.
The identification of such integrated, or pathwise, quantities \cite{spin16a} is generalized in our current formalism to read as follows:
\begin{minipage}{\linewidth}
\vspace{1em}
\noindent\rule[0.5ex]{\linewidth}{0.5pt}
\vspace{-2em}
\begin{definition}
By assuming the existence of unique measures, $\mathbb{P}_{X|\{Y\}}^{(s,r)}$ and $\mathbb{P}_{X}^{(s)}$, on a suitable path space for realizations of the stochastic process, $x_{t_0}^t$, we introduce the \emph{pathwise transfer entropy}
\begin{equation}
\mathcal{T}^{(s,r)}_{{y}\to{x}}[x_{t_0-s}^{t},y_{t_0-r}^{t}]=\ln{\frac{d\mathbb{P}_{X|\{Y\}}^{(s,r)}[x_{t_0}^{t}|x^{t_0}_{t_0-s},\{y^{t}_{t_0-r}\}]}{d\mathbb{P}_{X}^{(s)}[x_{t_0}^{t}|x^{t_0}_{t_0-s}]}}.
\label{local0}
\end{equation}
Equation (\ref{local0}) is a functional, mapping path functions of $x$ and $y$ into $\mathbb{R}$ ($\mathcal{T}_{y\to x}^{(s,r)}:\Omega_{xy}^{\mathbb{T}}\to\mathbb{R}$), designed to capture the information dynamics of individual realizations (path functions) of the stochastic processes where the measures are defined as those which satisfy the following property
\begin{equation}
T_{y\to x}^{(s,r)}\big|_{t_0}^t=\int_{t_0}^{t}\dot{T}^{(s,r)}_{{y}\to{x}}(t')dt'=\mathbb{E}_{P}\left[\mathcal{T}^{(s,r)}_{{y}\to{x}}[x_{t_0-s}^{t},y_{t_0-r}^{t}]\right].
\label{avint}
\end{equation}
\label{pathdef}
\end{definition}
\vspace{-2em}
\noindent\rule[0.5ex]{\linewidth}{0.5pt}
\end{minipage}
This should be interpreted as the continuous time generalization of Eqs.~(\ref{measure})$-$(\ref{localPathDiscrete}).
The contents of Eq.~(\ref{avint}) should be considered to be the total transfer entropy accumulated, or transferred, on the interval $[t_0,t)$.
We emphasize that this quantity is the expectation of the pathwise transfer entropy on the same time interval. This idea very closely resembles the concepts involved in modern treatments of entropy production, heat, work, etc., within formalisms such as stochastic thermodynamics \cite{seifert_stochastic_2008,seifert_stochastic_2012,spinney_fluctuation_2013}.

This leads to a dual definition of the transfer entropy rate, valid for stationary processes.
\begin{minipage}{\linewidth}
\vspace{1em}
\noindent\rule[0.5ex]{\linewidth}{0.5pt}
\vspace{-2em}
\begin{corol}
For stationary processes, such that $\dot{T}^{(s,r)}_{{y}\to{x}}$ is constant, Eq.~(\ref{avint}) implies
\begin{align}
\dot{T}^{(s,r)}_{{y}\to{x}}  &=\frac{1}{(t-t_0)}\mathbb{E}_{P}\left[\ln{\frac{d\mathbb{P}_{X|\{Y\}}^{(s,r)}[x_{t_0}^{t}|x^{t_0}_{t_0-s},\{y^{t}_{t_0-r}\}]}{d\mathbb{P}_{X}^{(s)}[x_{t_0}^{t}|x^{t_0}_{t_0-s}]}}\right].
 \label{eq:rateIntegralRelationshipStationary}
\end{align}
\label{pathwise}
\end{corol}
\vspace{-2em}
\noindent\rule[0.5ex]{\linewidth}{0.5pt}
\end{minipage}

The natural measures $\mathbb{P}_{X}^{(s)}$ and $\mathbb{P}_{X|\{Y\}}^{(s,r)}$ are those which jointly satisfy Eqs.~(\ref{eq:eq2}), (\ref{local0}), and (\ref{avint}) and also, along with an appropriate choice of path space, lead to the correct path properties in $x$ and may be understood as appropriate regular conditional probabilities of the measures $\mathbb{P}_{X}^{\mathbb{T},(s)}$ and $\mathbb{P}_{X|\{Y\}}^{\mathbb{T},(s,r)}$ on $(\Omega_x^{\mathbb{T}},\mathcal{F}_x^{\mathbb{T}})$. Identification of such measures will be implementation specific, but to satisfy the above we may state certain conditions on the finite dimensional distributions of the measures outlined in Appendix \ref{appAA}.

 We note that in the limit $s\to\infty$ we recover the canonical pushforward measure
  \begin{equation}
 \lim_{s\to\infty}\mathbb{P}_{X}^{\mathbb{T},(s)}[x_{\mathbb{T}}\in\mathcal{A}]=\mathbb{P}_{X}^\mathbb{T}[x_{\mathbb{T}}\in\mathcal{A}].
 \end{equation}
However, we emphasize
\begin{align}
&\lim_{\substack{s\to\infty\\r\to\infty}}\mathbb{P}_{X|\{Y\}}^{\mathbb{T},(s,r)}[x_{\mathbb{T}}\in \mathcal{A}|\{y_{\mathbb{T}}\}]\neq\mathbb{P}_{X|Y}^{\mathbb{T}}[x_{\mathbb{T}}\in \mathcal{A}|y_{\mathbb{T}}],
\end{align}
where the latter quantity is a conditional probability measure in the usual sense,
 i.e., the structure of $\mathbb{P}_{X|\{Y\}}^{\mathbb{T},(s,r)}$ [see also eq.~(\ref{meas})] does \emph{not} result in the standard definition of conditioning upon $y_{\mathbb{T}}$ because no details of the distributions of $y$ are included in its construction. We denote this distinction with the use of $\{\ldots\}$. This may be simultaneously thought of as the assumption that $y$ does not depend on $x$ or a recasting of the conditional dynamics into time inhomogeneous (non-stationary) dynamics parametrized by $y$. The distinction is most clearly described for a discrete time ($\mathbb{Z}\supseteq\mathbb{T}=[0,n]$) joint Markov process on a finite state space where one has $\mathbb{P}_{X|Y}^{\mathbb{T}}(x_0^n|y_0^n)=\mathbb{P}_{X,Y}^{\mathbb{T}}(x_0^n,y_0^n)/\mathbb{P}_{Y}^{\mathbb{T}}(y_0^n)=\mathbb{P}_0(x_0|y_0)\prod_{i=1}^{n}\mathbb{P}_{i}(x_i,y_i|x_{i-1},y_{i-1})/\mathbb{P}_{i}(y_i|y^{i-1}_0)$ whereas $\mathbb{P}_{X|\{Y\}}^{\mathbb{T}}(x_0^n|\{y_0^n\})=\mathbb{P}_0(x_0|y_0)\prod_{i=1}^{n}\mathbb{P}_i(x_i|x_{i-1},y_{i-1})$. The two expressions are only equivalent when $\mathbb{P}_{i}(x_i|x_{i-1},y_{i-1})=\mathbb{P}_{i}(x_i,y_i|x_{i-1},y_{i-1})/\mathbb{P}_{i}(y_i|y^{i-1}_0)$ which only happens in any generality when $y$ is independent of $x$ such that $\mathbb{P}_{i}(x_i,y_i|x_{i-1},y_{i-1})=\mathbb{P}_{i}(x_i|x_{i-1},y_{i-1}^i)\mathbb{P}_i(y_i|y_{i-1})$ and $\mathbb{P}_i(y_i|y^{i-1}_0)=\mathbb{P}_i(y_i|y_{i-1})$ and the transitions in $x$ and $y$ are not correlated such that $\mathbb{P}_{i}(x_i|x_{i-1},y_{i-1}^i)=\mathbb{P}_{i}(x_i|x_{i-1},y_{i-1})$. It cannot be overstated that $\mathbb{P}^{\mathbb{T}}_{X|\{Y\}}$ and $\mathbb{P}^{\mathbb{T}}_{X|Y}$ are distinct probability measures on $(\Omega^{\mathbb{T}}_x,\mathcal{F}^{\mathbb{T}}_x)$.
 
We also note that the approach for transfer entropy as a log-likelihood ratio for discrete time real-valued processes in \cite{barn12} is a special case of the general formalism for the pathwise transfer entropy in continuous time in Eq.~(\ref{local0}).\\
\\
Recent developments have discussed the importance of \emph{local} transfer entropy that is associated with individual transitions \cite{liz08a} [c.f. \eq{defLocal}]. We emphasize that the information dynamics of individual realizations here is captured by the pathwise transfer entropy and that any attempt to define a local transfer entropy rate may not be well defined. This is because the logarithm of the relevant Radon-Nikodym derivative may be non-differentiable and indeed may even be nowhere differentiable leading us to assert that a local transfer entropy rate may not exist\footnote{The smoothness of the pathwise transfer entropy is expected to follow that of paths $x_{\mathbb{T}}$. Consequently, it is expected that if these sample paths are non differentiable, such a quantity will not exist. This is the case in, for example, processes driven by Wiener noise or those that possess discontinuities. On the other hand such a quantity may exist for processes that emerge from a coarse graining of ordinary differential equations with smooth solutions.}:
\begin{minipage}{\linewidth}
\vspace{1em}
\noindent\rule[0.5ex]{\linewidth}{0.5pt}
\vspace{-2em}
\begin{prop}
A local or pointwise transfer entropy rate defined as
\begin{equation}
\dot{\mathcal{T}}^{(s,r)}_{y\to x}(t)=\lim_{dt\to 0}\frac{1}{dt}\ln{\frac{d\mathbb{P}_{t+dt}[{x}_{t+dt}|{{x}}^{t}_{t-s},{{y}}^{t}_{t-r}]}{d\mathbb{P}_{t+dt}[{x}_{t+dt}|{{x}}^{t}_{t-s}]}}
\end{equation}
 cannot be guaranteed to exist.
 \label{nolocal}
\end{prop}
\noindent\rule[0.5ex]{\linewidth}{0.5pt}
\end{minipage}

We finish by noting that all of the measure-theoretic and continuous time formalisms presented here are trivially extendible to conditioning on another source, or set of sources, to provide forms for the conditional transfer entropy \cite{liz08a,liz10e,verd05,vak09}.
\subsection{Implications for empirical work based on time discretization}
\label{sec:implicationsForEmpirical}
The overwhelming majority of the applications of transfer entropy in the literature concern empirical data from some real world process. Such underlying processes, despite being in continuous time, are often, in practice, sampled at a finite rate. Our main observation is the following:
\begin{minipage}{\linewidth}
\vspace{1em}
\noindent\rule[0.5ex]{\linewidth}{0.5pt}
\vspace{-2em}
\begin{remark}
\label{rem:discretisationLimit}
We recover an approximation to the quantities in this formalism given a discretization of a continuous time process by recognizing, due to the linearity of the expectation operator,
\begin{align}
&T^{(s,r)}_{{y}\to{x}}\big|_{t_0}^t=\nonumber\\
&\lim_{\Delta t\to 0}\mathbb{E}_{P}\left[\sum_{i=\lfloor t_0/\Delta t\rfloor+1}^{\lfloor t/\Delta t\rfloor}\ln{\frac{d\mathbb{P}_{i\Delta t}({x}_{i\Delta t}|{{x}}^{(i-1)\Delta t}_{(i-k)\Delta t},{{y}}^{(i-1)\Delta t}_{(i-l)\Delta t})}{d\mathbb{P}_{i\Delta t}({x}_{i\Delta t}|{{x}}^{(i-1)\Delta t}_{(i-k)\Delta t})}}\right]\nonumber\\
&=\lim_{\Delta t\to 0}\sum_{i=\lfloor t_0/\Delta t\rfloor+1}^{\lfloor t/\Delta t\rfloor}T^{(k,l)}_{y\to x}\big|_{(i-1)\Delta t}^{i\Delta t},\nonumber\\
k&=\left\lfloor\frac{s}{\Delta t}\right\rfloor+1, l=\left\lfloor\frac{r}{\Delta t}\right\rfloor+1
\label{eq:emp1},
\end{align}
where this limit exists, such that the relevant path measures are convergent in such a procedure, and where $\Delta t$ defines the discretization scheme.

Consequently the transfer entropy rate, given discretization of a continuous time process, would be approximated by
\begin{equation}
\dot{T}^{(s,r)}_{{y}\to{x}}(t)=\lim_{\Delta t\to 0}\frac{1}{\Delta t}T^{(k,l)}_{y\to x}\big|^{t}_{t-\Delta t},
\label{emp2}
\end{equation}
in line with eq.~(\ref{eq:eq2}).
\end{remark}
\noindent\rule[0.5ex]{\linewidth}{0.5pt}
\end{minipage}

Typical empirical assumptions and their implications are captured by the following:
\begin{minipage}{\linewidth}
\vspace{1em}
\noindent\rule[0.5ex]{\linewidth}{0.5pt}
\vspace{-2em}
\begin{remark}
When the process is both stationary and self averaging (ergodic), the transfer entropy rate would be estimated, in practice, by approximating the following limit:
\begin{align}
\dot{T}^{(s,r)}_{{y}\to{x}}&=\lim_{\substack{\Delta t\to 0\\(t-t_0)\to\infty}}\frac{1}{(t-t_0)}\nonumber\\
&\qquad\times\sum_{i=\lfloor t_0/\Delta t\rfloor+1}^{\lfloor t/\Delta t\rfloor}\ln{\frac{d\mathbb{P}_{i\Delta t}({x}_{i\Delta t}|{{x}}^{(i-1)\Delta t}_{(i-k)\Delta t},{{y}}^{(i-1)\Delta t}_{(i-l)\Delta t})}{d\mathbb{P}_{i\Delta t}({x}_{i\Delta t}|{{x}}^{(i-1)\Delta t}_{(i-k)\Delta t})}}\nonumber\\
&=\lim_{\substack{\Delta t\to 0\\(t-t_0)\to\infty}} \frac{1}{(t-t_0)}\nonumber\\
&\qquad\times\mathcal{T}_{{y}\to{x}}^{(k,l)}(x_{n\Delta t}^{(n+m)\Delta t},x_{(n-k)\Delta t}^{(n-1)\Delta t},y_{(n-l)\Delta t}^{(n-1+m)\Delta t})
\end{align}
with $n-1=\lfloor t_0/\Delta t\rfloor $, $n+m=\lfloor t/\Delta t\rfloor$, $k=\lfloor s/\Delta t\rfloor +1$, $l=\lfloor r/\Delta t\rfloor +1$, and where $\mathcal{T}_{{y}\to{x}}^{(k,l)}$ is accumulated over $m+1$ time steps as per eq.~(\ref{localPathDiscrete}).

\end{remark}
\noindent\rule[0.5ex]{\linewidth}{0.5pt}
\end{minipage}

\Eq{emp1} is consistent with the idea that one could, in principle, treat transfer entropy in continuous time as the limit of a discrete time transfer entropy and thus eq.~(\ref{eq:eq2}) as a discrete time transfer entropy rate as per eq.~(\ref{emp2}). We note, however, that the leading $\Delta t^{-1}$ term in eq.~(\ref{emp2}) has generally been overlooked (e.g. in \cite{ito11a}, where $T^{(k,l)}_{y\to x}\big|_{(i-1)\Delta t}^{i\Delta t}$ is computed for small $\Delta t$, but without the limit and the $\Delta t^{-1}$ term). 
This suggests that, where the limiting rate exists, a necessary condition for the appropriateness of the time-scale $\Delta t$ for a discrete time transfer entropy (in terms of capturing the time-scale of interactions, and not being undersampled) is that it must scale with $\Delta t$ in this vicinity.
We know for example that a limiting rate exists for linearly coupled Gaussian processes (with Wiener noise) in continuous time, where the Granger causality (proportional to transfer entropy for such processes \cite{barn09}) is linearly proportional to $\Delta t$ as $\Delta t \rightarrow 0$ \cite{barn16a,zhou14a}.

Furthermore, the above highlights a subtle distinction between transfer entropy as a statistic associated with a single instant in time as is common in the literature, and our interpretation which insists, even in discrete time, that transfer entropy can only ever be associated with an accumulation over a finite time interval even if that interval is simply one time step. In contrast it is the transfer entropy rate that exists for instances in time. In other words, in discrete time, if each time step is considered to take to a value of one, but is otherwise \emph{dimensionless}, we have $T_{y\to x}^{(k,l)}(n)=T_{y\to x}^{(k,l)}\big|^n_{n-1}=\Delta_n{T}_{y\to x}^{(k,l)}(n)$ (where $\Delta_n$ indicates a discrete time derivative on $\mathbb{Z}$ analogous to the usual time derivative on $\mathbb{R}$)\footnote{We note that for stationary processes, this generalizes to $m^{-1}T_{y\to x}^{(k,l)}\big|^{n+m}_{n}$.}. However, as soon as one associates some unit or dimension with time one is obliged to distinguish between those quantities in nats (or bits) and those in nats per unit time. If each time step is deemed, still, to take value one, the quantities, while distinct, have the same value, leading to the previously discussed ambiguity. But, application to continuous time shows that in general these notions are distinct and we argue that one should always, in continuous or discrete time, whether time is physical or otherwise, distinguish between accumulated transfer entropies (in nats), which can only exist on a finite time interval, and transfer entropy rates (in nats per unit time).

Finally, we note that the approach in Remark \ref{rem:discretisationLimit} unavoidably leads to a divergence in the number of bins required to capture path histories which we expect to be seriously limiting in practice.\footnote{For the simplest state spaces, $\Sigma_x$ and $\Sigma_y$ being binary, the full sample space required for the calculation would be $2^{1+{s/{\Delta t}}+{r/{\Delta t}}}$. A relevant example here is of neural spike trains, where a typically relevant path history would be of order 200 ms (see e.g. \cite{ald91,bargad01}) in both source and target, at a conservative 1 ms interaction resolution (noting that finer resolution would be more desirable), meaning that a naive discrete implementation would explore a state space of $2^{200+200+1}$ potential configurations. The number of samples and thus the time and memory requirements for estimation scales at least on this order, and therefore becomes impractical.} While this may seem unpromising for real world applications outside of theoretical models where path measures can either be asserted or derived, there do exist classes of stochastic processes, in continuous time, where alternative representations exist such that no binning is required. Where real world phenomena can be  meaningfully approximated by such stochastic processes we can then dramatically improve this picture. Such processes are the subject of the next section.

\section{Jump processes}
\label{sec:jump}
For the remainder of this paper we now focus specifically on jump processes. These are stochastic processes characterized by intermittent transitions between states in $\Sigma_x$ and where the states are constant in-between these transitions. They can be thought of as a non-Markov, inhomogeneous and possibly non-stationary generalization of compound-Poisson or renewal-reward processes. As such we consider $\Omega^{\mathbb{T}}_x$ to be the space of \emph{c\`adl\`ag} (right continuous with left limits) step functions on $\Sigma_x$ (therefore $\mathcal{F}^{\mathbb{T}}_x$ is taken to be the Borel sigma algebra associated with the $J_1$, or Skorokhod, topology on $\Omega^{\mathbb{T}}_x$ \cite{billingsley_convergence_1968}). We note that we present a formalism for discrete state spaces, $\Sigma_x$, with the power set $\mathcal{X}=2^{\Sigma_x}$, which necessarily deal with summations over states, but this is trivially modified for use with continuous state spaces by replacing all sums by the appropriate integrals (or indeed more complicated spaces by an integral w.r.t an appropriate measure). Examples of such systems are ubiquitous, but include financial times series such as equity prices, population dynamics, and spiking neural processes.
\begin{minipage}{\linewidth}
\vspace{1em}
\noindent\rule[0.5ex]{\linewidth}{0.5pt}
\vspace{-2em}
\begin{prop}
For stochastic processes $\{x_t\}_{t\in \mathbb{T}}$, $\mathbb{T}\subseteq\mathbb{R}$, whose sample paths are \emph{c\`adl\`ag} step functions which permit description by transition rates $W$ and escape rates $\lambda$, with path $x_{t_0}^{t}$ captured by the starting configuration $x_0$ at time $t_0$, $N_x$ transitions into states $x_i$ at times $t_i$ up until final time $t$, the \emph{pathwise transfer entropy} is given by \cite{spin16a}:
\begin{align}
\mathcal{T}^{(s,r)}_{{y}\to{x}}[x_{t_0}^{t}\equiv&\{t,\{t,x\}_0^{N_x}\},y_{t_0}^{t}]=\sum_{i=1}^{N_x}\ln{\frac{W[x_{i}|x^{t_i}_{{t_{i}}-s},y^{t_i}_{{t_{i}}-r}]}{W[x_{i}|x^{t_i}_{{t_{i}}-s}]}}\nonumber\\
&+\int_{t_0}^{t}\left(\lambda_{x}[{{x}}_{t'-s}^{t'}]-\lambda_{x|y}[{{x}}_{t'-s}^{t'},{{y}}_{t'-r}^{t'}]\right)dt',
\label{eq:local}
\end{align}
\label{pathjump}
where $\{t,x\}_0^{N_x}\equiv\{\{t_0,x_0\},\ldots,\{t_N,x_N\}\}$ indicates the set of states and times which, in addition to the final time $t$, defines the path.
\end{prop}
\vspace{-1em}
\noindent\rule[0.5ex]{\linewidth}{0.5pt}
\end{minipage}
Intuitively, the origin of the distinct terms in Eq.~(\ref{eq:local}) may be understood as a summation of terms that correspond to the ``surprise'' of observing transitions to $x_i$ at times $t_i$ plus the continuous limit of a summation of surprise contributions arising from non-transitioning behavior.
	
To present the above, we begin by formally defining our notation.
In such systems the quantities which characterize the behavior are transition rates, for which we require those with and without knowledge of the source $y$. We may construct them, using the probability of the $\mathcal{F}^{\mathbb{T}}_x$-measurable event of having a transition in a given interval $[a,b]$ denoted here by $\mathbb{P}_{[a,b]}$, by writing
\begin{align}
W[{x}'|{{x}}_{t-s}^{t}&,{{y}}_{t-r}^{t}]\nonumber\\
=\lim_{dt\to 0} &\frac{1}{dt}\mathbb{P}_{[t,t+dt]}[{x}'\in[t,t+dt]|{{x}}_{t-s}^{t},{{y}}_{t-r}^{t}],\nonumber\\
W[{x}'|{{x}}_{t-s}^{t}]&=\lim_{dt\to 0} \frac{1}{dt}\mathbb{P}_{[t,t+dt]}[{x}'\in[t,t+dt]|{{x}}_{t-s}^{t}],
\label{Wrates}
\end{align}
where the notation ${x}'\in[t,t+dt]$ indicates the transition into state ${x}={x}'$ in the interval $[t,t+dt]$. This naturally leads to the mean escape rates
\begin{align}
&\lambda_{x}[{{x}}_{t-s}^{t}]=\sum_{{x}'\neq {x}^{-}_t}W[{x}'|{{x}}_{t-s}^{t}],\\
&\lambda_{x|y}[{{x}}_{t-s}^{t},{{y}}_{t-r}^{t}]=\sum_{{x}'\neq {x}^{-}_t}W[{x}'|{{x}}_{t-s}^{t},{{y}}_{t-r}^{t}],
\end{align}
which are the rates of transitioning out of state $x_t$, given knowledge of the history of $x$ or both $x$ and $y$ and where $x_t^{-}=\lim_{t'\nearrow t}x(t')$. We have made no assumption about the nature of $y$, however, if $y$ is also a jump process on a discrete state space we have $W[{x}'|{{x}}_{t-s}^{t},{{y}}_{t-r}^{t}]=\sum_{y'}W[{x}',{y}'|{{x}}_{t-s}^{t},{{y}}_{t-r}^{t}]$.  We note that such processes do not possess an embedded discrete time process such as an embedded Markov chain since we consider non-Markovian potentially non-stationary processes. Again we point out we recover Markovian transition and escape rates in the limit $s\searrow 0$, $r\searrow 0$.\\
\\

In Appendix \ref{appA}, we use the above quantities to construct the relevant probability measures of a jump process, $x_{t_0}^t$, running from time $t'=t_0$ to time $t'=t$ that are consistent with the relevant finite dimensional distributions [Eqs.~(\ref{PX}) and (\ref{meas})]. We introduce notation such that for a path that consists of $N_x$ transitions in $x$, transitions may be labeled by the index $i\in\{1,\ldots,N_x\}$ so that $x_i\equiv x_{t_i}$ being the state into which the system transitions at time $t_i$. 
We maintain the notation for the initial time, $t_0$, and introduce notation for the initial state $x_0=x^-_{t_0}$ to exploit the indexing system as a deliberate abuse of notation to characterize the path up to the first transition. Key results include the identification of the following probability densities (which may be thought of as generalized Janossy densities \cite{daley_introduction_2003}) w.r.t. the Lebesgue measure on $\mathbb{R}^{N_x}$, or likelihoods, for a specific path realization arising from measures $\mathbb{P}_X^{(s)}$ and $\mathbb{P}_{X|\{Y\}}^{(s,r)}$, respectively,
\begin{align}
&p_{N_x}^{(s)}[x^t_{t_0}\equiv\{t,\{t,x\}_0^{N_x}\}|x^{t_0}_{t_0-s}]\nonumber\\
&\quad=\left(\prod_{i=1}^{N_x}W[x_{i}|x^{t_i}_{{t_{i}}-s}]\right)\exp{\left[-\int_{t_0}^{t}\lambda_x[x^{t'}_{t'-s}]dt'\right]},
\label{eq:probden0-maintext}
\end{align}
\begin{align}
&p_{N_x}^{(s,r)}[x^t_{t_0}\equiv\{t,\{t,x\}_0^{N_x}\}|x^{t_0}_{t_0-s},\{y^{t}_{t_0-r}\}]\nonumber\\
&\qquad\qquad=\left(\prod_{i=1}^{N_x}W[x_{i}|x^{t_i}_{{t_{i}}-s},y^{t_i}_{{t_{i}}-r}]\right)\nonumber\\
&\qquad\qquad\quad\times\exp{\left[-\int_{t_0}^{t}\lambda_{x|y}[x^{t'}_{t'-s},y^{t'}_{t'-r}]dt'\right]}.
\label{probden-maintext}
\end{align}
We note $\{t,x\}_0^{N_x}=\{\{t_0,x_0\}\dots\{t_{N_x},x_{N_x}\}\}$ such that we can represent any path $x^t_{t_0}\equiv\{t,\{t,x\}_0^{N_x}\}$. We point out that expectations are taken w.r.t. these measures by implementing variants of the following infinite series for $\mathbb{P}_{X}^{(s)}$
\begin{equation}
\mathbb{E}_{\mathbb{P}_{X}^{(s)}}\left[f[x^t_{t_0}]\right]=\int_{\Omega_x}f[x_{t_0}^t]d\mathbb{P}_{X}^{(s)}[x_{t_0}^t]=\sum_{i=0}^{\infty}J^f_i[t,x^{t_0}_{t_0-s}]
\label{Jsum}
\end{equation}
where
\begin{widetext}
\begin{align}
&J^f_i[t,x^{t_0}_{t_0-s}]=\sum_{\substack{x_1\in\Sigma_{x} \\ x_1\neq x_{0}}}\ldots \sum_{\substack{x_{i}\in\Sigma_{x} \\ x_{i}\neq x_{i-1}}}\int_{t_0}^{t}dt_1\ldots\int_{t_{i-1}}^{t}dt_{i}f_{i}(x^t_{t_0}\equiv\{t,\{t,x\}_0^{i}\}) p_{i}^{(s)}[x^t_{t_0}\equiv\{t,\{t,x\}_0^{i}\}|x^{t_0}_{t_0-s}]
\end{align}
\end{widetext}
and where
\begin{align}
&J^f_0[t,x^{t_0}_{t_0-s}]=\nonumber\\
&f_{0}(x^t_{t_0}\equiv\{t,\{t_0,x_0\}\})p_0^{(s)}[x^t_{t_0}\equiv\{t,\{t_0,x_0\}\}|x^{t_0}_{t_0-s}].
\end{align}
Here $f_i$ are the functional forms of $f$ given $i$ transitions in $x$. When $f[x_{t_0}^t]=1$ we have $J_i^1[t,x^{t_0}_{t_0-s}]$ equal to the probabilities of having $i$ transitions on $[t_0,t)$, given $x^{t_0}_{t_0-s}$, such that $\sum_{i=0}^{\infty}J_i^1[t,x^{t_0}_{t_0-s}]=1$.
Explicitly, $p_i^{(s)}$ is the probability density for a path on $t'\in[t_0,t)$ that contains $i$ transitions, conditional upon the previous path function $x_{t_0-s}^{t_0}$, where transition rates utilize $s$ seconds of history dependence. We note that $p_i^{(s)}$ would also be a density with respect to $\{x_1,\ldots,x_{N_x}\}$ should $x$ be continuous. 

Given such quantities, identified in Appendix \ref{appA}, the Radon Nikodym derivative may be identified as the ratio of such probability densities, or log likelihood ratio \cite{billingsley_probability_1995}, and thus the pathwise transfer entropy in eq.~(\ref{local0}) as the sum and integral contribution in \eq{local} appearing in Proposition \ref{pathjump}.

Explicitly \eq{local}, the pathwise transfer entropy, consists of:
\begin{enumerate}
 \item a continuously varying contribution (associated with the waiting times between transitions), that is interrupted by
 \item discontinuous jump contributions arising when a transition in $x$ occurs.
\end{enumerate}
In both cases the terms can be interpreted as arising from differences in surprisal, but from the distinct nontransitioning and transitioning behavior along the path. The implication is that not only can a transition be predicted by the previous behavior in $x$ and $y$, but the absence of a transition can as well.\\
\\
Examining the pathwise transfer entropy in \eq{local}, we can consider analogs to the local or pointwise contributions associated with the usual formalism of transfer entropy \cite{liz08a} by considering the contributions associated with transitions and periods between them. Doing so allows us to consider a \emph{local} contribution to the transfer entropy associated with a transition $\Delta \mathcal{T}^{(s,r)}_{t}(t_i)$ and a \emph{local} rate of transfer entropy associated with periods in-between transitions $\dot{\mathcal{T}}_{nt}^{(s,r)}(t)$ such that
\begin{align}
&\mathcal{T}^{(s,r)}_{{y}\to{x}}[x_{t_0}^{t}\equiv\{t,\{t,x\}_0^{N_x}\},y_{t_0}^{t}]\nonumber\\
&\qquad=\sum_{i=1}^{N_x}\Delta \mathcal{T}^{(s,r)}_{t}(t_i)+\int_{t_0}^{t}\dot{\mathcal{T}}_{nt}^{(s,r)}(t')dt',
\label{eq:localContributions}
\end{align}
with $\Delta \mathcal{T}^{(s,r)}_{t}(t_i)$ and $\dot{\mathcal{T}}_{nt}^{(s,r)}(t)$ defined by identification with \eq{local}. However, we point out that these two contributions are distinct, and any attempt to produce a single \emph{local} (pointwise) \emph{rate} will be rendered divergent because of the discontinuous contributions at the transitions, thus confirming Proposition \ref{nolocal}. 

Next we consider the (average) transfer entropy rate for jump processes:
\begin{minipage}{\linewidth}
\vspace{1em}
\noindent\rule[0.5ex]{\linewidth}{0.5pt}
\vspace{-2em}
\begin{prop}
The \emph{transfer entropy rate} for jump processes, as described, is given by the expectation
\begin{align}
\dot{T}^{(s,r)}_{{y}\to{x}}(t)&=\mathbb{E}_{P}\left[(1-\delta_{x_t^- x_t})\ln{\frac{W[x_{t}|x^{t}_{{t}-s},y^{t}_{{t}-r}]}{W[x_{t}|x^{t}_{{t}-s}]}}\right]
\label{av}
\end{align}
where $\delta_{x_t^- x_t}$ is the Kronecker delta function.
\label{ratejump}
\end{prop}
\noindent\rule[0.5ex]{\linewidth}{0.5pt}
\end{minipage}
Crucially, the \emph{expectation} of the contribution to the transfer entropy rate associated with non-transitioning behavior vanishes. This arises directly from the property
\begin{equation}
\mathbb{E}_{P}[\lambda_{x|y}[{{x}}_{t-s}^{t},{{y}}_{t-r}^{t}]]=\mathbb{E}_{P}[\lambda_{x}[{{x}}_{t-s}^{t}]]=\mathbb{E}_{P}[\lambda_{x\cdot}[\cdot]]
\end{equation}
since each is simply an expression for the mean escape rate in $x$, achieved by averaging over all relevant path histories. This is naturally independent of the details of such histories since each expression is a linear sum of transition rates which can be directly marginalized.
Consequently, by exchanging the order of the expectation and integral, we have
\begin{equation}
\mathbb{E}_{P}\left[\int_{t_0}^{t}\left(\lambda_{x}[{{x}}_{t'-s}^{t'}]-\lambda_{x|y}[{{x}}_{t'-s}^{t'},{{y}}_{t'-r}^{t'}]\right)dt'\right]=0
\label{nonspike}
\end{equation}
and thus $\mathbb{E}_{P}[\dot{\mathcal{T}}_{nt}^{(s,r)}(t)]=0$.
As such, there is no net contribution to the expected rate arising from the pathwise transfer entropy associated with waiting times between target transitions.
Consequently, the transfer entropy rate is expressible by Eq.~(\ref{av}) in Proposition \ref{ratejump}.
We point out that such an expectation is computed in a similar manner to Eq.~(\ref{Jsum}) where, in this instance, we have $\mathbb{E}_P=\mathbb{E}_{\mathbb{P}_{XY}^{\mathbb{T}}}$, but with all permutations of transitions in $x$ and $y$ as opposed to just in $x$. For instance, if $f=1$ we would have $\sum_{i=0}^{\infty}\sum_{j=0}^{\infty}J_{i,j}^1=1$ where $i$ and $j$ are the number of transitions in $x$ and $y$, respectively. We also point out that in each $J_{i,j}$ term the leading $(1-\delta_{x_{t}^{-}x_{t}})$ manifests as a Dirac delta $\delta(t_i-t)$, where $t_i$ is the $i$th transition in $x$, with units $t^{-1}$, confirming the expression is dimensionally sound.

Again, we compare this to the implied empirical formulation for a self-averaging stationary process which can be expressed through the following:
\begin{minipage}{\linewidth}
\vspace{1em}
\noindent\rule[0.5ex]{\linewidth}{0.5pt}
\vspace{-2em}
\begin{remark}
For stationary self averaging processes the \emph{transfer entropy rate} is equivalent to the implied empirical measurement strategy
\begin{equation}
\dot{T}^{(s,r)}_{{y}\to{x}}=\lim_{(t-t_0)\to\infty}\frac{1}{(t-t_0)}\sum_{i=1}^{N_x}\ln{\frac{W[x_{t_i}|x^{t_i}_{{t_{i}}-s},y^{t_i}_{{t_{i}}-r}]}{W[x_{t_i}|x^{t_i}_{{t_{i}}-s}]}}
\label{empjump}
\end{equation}
where $N_x$ is the number of transitions in $x$ in the interval $[t_0,t)$.
\label{empjumprem}
\end{remark}
\noindent\rule[0.5ex]{\linewidth}{0.5pt}
\end{minipage}
Crucially we can see that in comparison to \eq{emp1}, no limit in a time discretization parameter is required; Eq.~(\ref{empjump}) is asymptotically exact as $t\to\infty$ which may be achieved empirically by simply considering more data. 
Finally, as per \secRef{contTime}, all of the formalisms for jump processes are trivially extendible to conditional transfer entropies \cite{liz08a,liz10e,verd05,vak09}.

\section{Application to spike trains}
\label{sec:spikes}

Next, we turn our attention to point processes, the most prominent example of which being spike train processes common to neuroscience. These processes are not characterized by transitions between distinct states, but rather consist of path spaces which permit, in model, several non-overlapping and individually indistinguishable events or spikes of zero width which occur in continuous time. As such, the paths are completely described by the times of such spikes. To apply the preceding formalism, we must consider them as a \emph{c\`adl\`ag} process with the most natural way being to recast them as a non-Markov extension of a Poisson counting process or a generalized modulated renewal process, which in turn may be multidimensional. In such a setup, the spike rate is equivalent to the rate of increasing the counting process by one or the transition rate between ``state'' $N$ and $N+1$ where $N$ is the total number of spikes that have occurred. Here, $N$ is arbitrary and so we insist that any transition rate be independent of $N$ such that the \emph{path dependent spike rate} (or conditional intensity function) is 
\begin{align}
W[x^t_{\text{spike}}|x^{t}_{t-s}]&=W[N^{-}_{t}+1|N^{t}_{t-s}]\nonumber\\
&=W[N^{-}_{t}+1+m|N^{t}_{t-s}+m]\quad\forall \:m\in\mathbb{N}
\end{align}
where $N^{t}_{t-s}+m$ indicates that $m$ spikes have been uniformly added to the counting process and $x^t_{\text{spike}}$ indicates a spike in $x$ at time $t$. Such a process, in state $N_t$ may only escape into state $N_t+1$ (i.e. not state $N_t+2$ etc.) meaning that we also recognize that 
\begin{equation}
W[N^{-}_{t}+1+m|N^{t}_{t-s}+m]=\lambda_x[N^{t}_{t-s}+m]=\lambda_x[x^{t}_{t-s}]
\end{equation}
such that the path dependent spike rates act as both the path dependent transition and escape rates. 
In the first instance, this simplifies eqs.~(\ref{eq:probden0-maintext}) and (\ref{probden-maintext}) (see also \cite{boss16a,solo07,okat05a,kim11a,Truccolo1074}).
 Returning, for continuity, to an expression of paths, $x$, we can represent any path containing $N_x$ spikes starting at time $t_0$ as $x^t_{t_0}\equiv\{t,\{t\}_0^{N_x}\}$ with spike times $\{t\}_0^{N_x}=\{t_0,\ldots,t_{N_x}\}$.
  
  By comparison with Eqs.~(\ref{eq:local}) and (\ref{av}) we then have
\begin{minipage}{\linewidth}
\vspace{1em}
\noindent\rule[0.5ex]{\linewidth}{0.5pt}
\vspace{-2em}
\begin{prop}
For spike train or point processes, the \emph{pathwise transfer entropy} is given by
\begin{align}
&\mathcal{T}^{(s,r)}_{{y}\to{x}}[x_{t_0}^{t}\equiv\{t,\{t\}_0^{N_x}\},y_{t_0}^{t}]=\sum_{i=1}^{N_x}\ln{\frac{\lambda_{x|y}[x^{t_i}_{{t_{i}}-s},y^{t_i}_{{t_{i}}-r}]}{\lambda_x[x^{t_i}_{{t_{i}}-s}]}}\nonumber\\
&\quad+\int_{t_0}^{t}\left(\lambda_{x}[{{x}}_{t'-s}^{t'}]-\lambda_{x|y}[{{x}}_{t'-s}^{t'},{{y}}_{t'-r}^{t'}]\right)dt'.
\label{spikepath}
\end{align}
\label{spikepathprop}
\end{prop}
\vspace{-1em}
\noindent\rule[0.5ex]{\linewidth}{0.5pt}
\begin{prop}
For spike train or point processes, the \emph{transfer entropy rate} is given by the expectation
\begin{align}
\dot{T}^{(s,r)}_{{y}\to{x}}(t)&=\mathbb{E}_{P}\left[(1-\delta_{x_t^- x_t})\ln{\frac{\lambda_{x|y}[x^{t}_{{t}-s},y^{t}_{{t}-r}]}{\lambda_{x}[x^{t}_{{t}-s}]}}\right].
\label{spikeav}
\end{align}
\label{spikeavprop}
\end{prop}
\vspace{-2em}
\noindent\rule[0.5ex]{\linewidth}{0.5pt}
\end{minipage}

These quantities have the same properties as the more general jump processes case.
That is, eq.~(\ref{spikepath}), the pathwise transfer entropy, consists of:
\begin{enumerate}
 \item a continuously varying contribution (relating to waiting times between spikes), with rate $\dot{\mathcal{T}}_{nt}^{(s,r)}(t)$; that is interrupted by
 \item discontinuous jump contributions, $\Delta \mathcal{T}^{(s,r)}_{t}(t_i)$, when a spike in $x$ occurs.
\end{enumerate}
Again, this implies that not only can a spike in the target $x$ be predicted by the previous behavior in $x$ and $y$, but the absence of a spike can as well. However, there is no net contribution to the expected rate arising from the pathwise transfer entropy associated with waiting times between target spikes.

The implied empirical formalism in this case, again for stationary self averaging processes, is of the form in Eq.~(\ref{empspike}) in Remark \ref{empspikerem} and thus reads
\begin{minipage}{\linewidth}
\vspace{1em}
\noindent\rule[0.5ex]{\linewidth}{0.5pt}
\vspace{-2em}
\begin{remark}
For stationary self averaging point processes the transfer entropy rate is equivalent to the implied empirical measurement strategy
\begin{equation}
\dot{T}^{(s,r)}_{{y}\to{x}}=\lim_{(t-t_0)\to\infty}\frac{1}{(t-t_0)}\sum_{i=1}^{N_x}\ln{\frac{\lambda_{x|y}[x^{t_i}_{{t_{i}}-s},y^{t_i}_{{t_{i}}-r}]}{\lambda_x[x^{t_i}_{{t_{i}}-s}]}}
\label{empspike}
\end{equation}
where $N_x$ is the number of spikes in $x$ in the interval $[t_0,t)$.
\label{empspikerem}
\end{remark}
\noindent\rule[0.5ex]{\linewidth}{0.5pt}
\end{minipage}
At this point we wish to point out that for such continuous time processes the ability to losslessly represent paths $x^t_{t_0}\equiv\{t,\{t\}_0^{N_x}\}$ points to a strategy for efficient empirical computation, as an alternative to brute force time discretization approaches, to be presented in a companion paper.

The idea that information in spike times relates to an underlying directed relationship has been observed, e.g., in \cite{strong98} and regarding ``causal entropy'' in \cite{zoch04a,wadd07a}, which indeed computed entropies of (cross) inter-spike intervals. However to our knowledge, this is the first formulation that computes transfer entropy based on lossless representation of entire spike trains (and is thus a dynamic quantity which captures state-updates rather than static correlations of single spike-time relationships).
We also note that our formulation would capture information transmission facilitated via either rate or temporal coding \cite{butts07a}.

We take a moment to point out that in order to describe a genuinely non-parametric statistic such as the transfer entropy, such a formalism must be completely general and so can easily capture the dynamics of frequently used processes for neural modeling. For instance such a formalism can represent a non-stationary Poisson process, $\lambda_x[x^t_{t-s}]=\lambda_x(g(t))$, a modulated renewal process, $\lambda_x[x^t_{t-s}]=\lambda_x(g(t),t-t_{N_x})$, where $g(t)$ is a time varying protocol with the same continuity properties as $x$, or higher order stochastic processes such as Cox processes through $\lambda_{x|y}[x^t_{t-s},y^t_{t-r}]=\lambda_{x|y}(y_t^{-})$ \cite{daley_introduction_2003}. Indeed, we assume some dependence on another variable in order for the concept of transfer entropy to be relevant. We emphasize, however, that the hidden variables used in the construction of such processes need not be the source used in the calculation of the transfer entropy (i.e., the doubly stochastic variable in a Cox process could be some hidden variable $z$, for instance). And indeed, such hidden variables (or others) could be trivially  conditioned on in all of these formalisms for spiking processes to make the extension to conditional transfer entropies as discussed in \secRef{jump} \cite{liz08a,liz10e,verd05,vak09}.

\section{Examples}
\label{sec:examples}
To highlight the properties of our results we present two examples of spike train processes where, analytically and numerically, respectively, the transfer entropy can be calculated. In these examples, both the target and source are considered to be point processes. We point out that for such spike train processes the transition rate in $x$ where $y$ is known must have some finite non-Markov character dependent on the history of $y$ since otherwise the process maps to the same Markovian Poisson process independently of the knowledge of $y$ giving a transfer entropy of zero. The main challenge for analytical computation is the tractability of computing the coarse grained spike rate $\lambda_x$ since, as mentioned above, the joint process must be non-Markov. 

\subsection{Simple analytical example}
In our first example, we alleviate such difficulties by defining a process and considering it in the regime where it is feasible to calculate the coarse grained spike rate analytically. To do so, we consider a simple model of neuron spiking. In this model, a source neuron spikes randomly with a refractory period preventing rapid sequential spiking. Source spikes can cause a target neuron, also with a refractory period, to spike with a defined probability within a subsequent time window. We can summarize the process with the following statements:
\begin{itemize}
\item Both the source $y$ and target $x$ each have independent refractory periods of duration $\tau^r$ following a spike, during which they cannot spike.
\item Outside of its refractory period, the source $y$ is a regular, stationary, and Markovian, Poisson process with rate $\lambda_y$ and is independent of $x$.
\item The target $x$ may spike only within a window of $\tau$ seconds duration following a spike in the source $y$. The probability of $x$ spiking in the interval is $a$. This leads to an elevated spike rate in the $\tau$ long interval of $\lambda_{x|y}^{e}=-\tau^{-1}\ln{[1-a]}$ since the probability of $x$ not spiking in this window is $e^{-\int_{0}^{\tau}\lambda_{x|y}^e dt}$.
\item The refractory period $\tau^r$ is longer or equal to the elevated rate period $\tau$, a by-product of which being that that the target $x$ may only spike once in the elevated rate period $\tau$.
\item The target and source cannot spike simultaneously. Such a property is sometimes called bipartite. This means $\lambda_{x|y}[x_{t-s}^{t},y_{t-r}^{t}]=W[x^t_{\text{spike}},y^{-}_t|x_{t-s}^{t},y_{t-r}^{t}]$.
\item Spike rates $\lambda_{x|y}[x_{t-s}^{t},y_{t-r}^{t}]$, as functionals of \emph{c\`adl\`ag} paths $x$ and $y$, are therefore defined at time $t$ with (up to) the left limit values of $x$ and $y$ and so themselves must be \emph{c\`agl\`ad} (left continuous with right limits) when viewed as functions of $t$.
\end{itemize}
We can summarize the above by representing the transition rates, in the limit $s,r \to \infty$, as $\lambda_{y|x}[x^t_{t-s},y^t_{t-r}]=\lambda_{y|x}(t,t^y)$ and $\lambda_{x|y}[x^t_{t-s},y^t_{t-r}]=\lambda_{x|y}(t,t^y,t^x)$ where $t^y<t$ and $t^x<t$ are the times of the most recent spikes in the source and target respectively, such that
\begin{align}
\lambda_{y|x}(t,t^y)&=\begin{cases}\lambda_y,&t> t^y+\tau^r\\
0,&t\leq t^y+\tau^r
\end{cases}
\nonumber\\
\lambda_{x|y}(t,t^y,t^x)&=\begin{cases}
\lambda_{x|y}^e=-\frac{1}{\tau}\ln{[1-a]}, & t^y < t\leq t^y+\tau\\
&t^x\leq t^y\\
&t> t^x+\tau^r\\
0, &  \text{otherwise}.
\end{cases}
\label{spikerates}
\end{align}
We then consider this process up to first order in $\lambda_y$. The critical step in computing relevant quantities (the transfer entropy rate and pathwise transfer entropy) is in approximating the coarse grained $\lambda_x$. In this regime, it can be shown [see Appendix \ref{appB} for a complete treatment in the $\mathcal{O}(\lambda_y)$ regime] that the coarse grained rate, as a function of the single most recent spike in $x$, is given by\footnote{Equation ~(\ref{eq:simpleExampleDependentRate}) is an estimate of the $\mathcal{O}(\lambda_y)$ coarse grained rate, $\lambda_x$, as a function of an arbitrary multi-spike history, but agrees when the interspike interval between the first and second most recent spikes in the arbitrary history is greater than $\tau^r+\tau$. This condition dominates the path histories in the $\mathcal{O}(\lambda_y)$ regime since paths with $N_x$ spikes have probability density with leading order terms $\mathcal{O}(\lambda_y^{N_x})$ since every spike in $x$ is preceded by one in $y$.}

\begin{align}
 &\lambda_x(t,t^x=0)\nonumber\\
 &=\begin{cases}
 0,&0\leq t<\tau^r\\
 \left(1-(1-a)^{\frac{t-\tau^r}{\tau}}\right)\lambda_y,&\tau^r\leq t<\tau^r+\tau\\
 a\lambda_y,&t\geq \tau^r+\tau
 \end{cases}\nonumber\\
 &\quad+\mathcal{O}(\lambda_y^2).
 \label{eq:simpleExampleDependentRate}
 \end{align}
We understand that, in this regime, from a perspective without knowledge of $y$, after any given spike the $x$ process appears to be described by a refractory period of duration $\tau^r$ as before, a subsequent period in which the spike rate grows, then a regime from $\tau+\tau^r$ seconds after a spike when it is readily approximated as a Markovian Poisson process with rate $a\lambda_y$. Such a form could then be readily used to calculate the pathwise transfer entropy using Eq.~(\ref{spikepath}).

\begin{figure}[h!]
\centering
\includegraphics[scale=0.6]{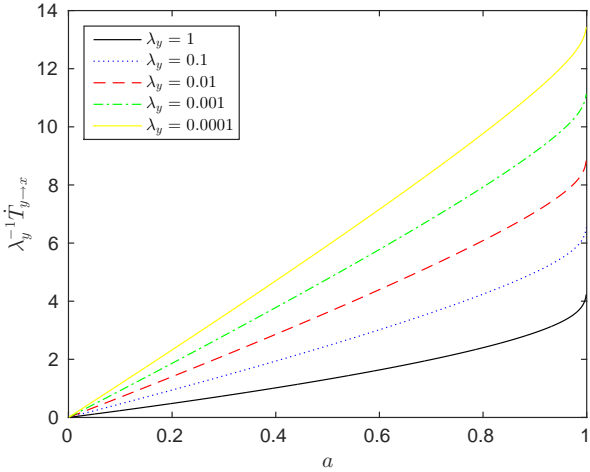}
\caption{\label{fig1}The transfer entropy rate (normalized by the limiting mean target spike rate) for the process obeying rates as described in Eq.~(\ref{spikerates}), with $\dot{T}_{y\to x}$ correct to $\mathcal{O}(\lambda_y)$, $\tau=1$.}
\end{figure}

The same spike rate can then be utilized to calculate the transfer entropy rate (a full treatment is found in Appendix \ref{appB}). Crucially, when performing the requisite path integral average, the relevant path probability density introduces an additional term in $\lambda_y$. Consequently, for this particular calculation in this regime, this has the effect of permitting us to exclude higher order terms associated with multiple spikes allowing for an even simpler approximation for $\lambda_x$, equivalent to considering it to be a Markov Poisson process with rate $a\lambda_y$ throughout. This yields, again with $s\to\infty$, $r\to\infty$,
\begin{equation}
\dot{T}_{{y}\to{x}}=a\lambda_y\ln{\left[\frac{-\ln{[1-a]}}{a\lambda_y\tau}\right]}+\mathcal{O}(\lambda_y^2).
\label{TErateexA}
\end{equation}
 The variation of the transfer entropy rate is shown in Fig.~(\ref{fig1}).
The form of eq.~(\ref{TErateexA}) reflects the fact that the appropriate approximation is equivalent to considering the spike process in $x$ to be a Markov Poisson process with rate $a\lambda_y$ when there is no knowledge of $y$ and a Markov Poisson with rate $\lambda_{x|y}^e=-\tau^{-1}\ln{[1-a]}$ when $y$ is known. Small increases in $\lambda_y$ lead to increases in the transfer entropy rate, but only because of the subsequent increase in the likelihood of a spike in $x$ reflected in the leading $a\lambda_y$ term. On the other hand, we observe a decrease with the same small increase in $\lambda_y$ in the transfer entropy rate normalized by this limiting mean target spike rate, since the increased likelihood renders each spike less surprising and thus less informative.
Further, as $a$ increases, the transfer entropy also increases because the predictability of $x$ with knowledge of $y$ increases since one can be increasingly confident that a spike in $x$ will occur. When $a\to 1$ or $\tau\to 0$, the transfer entropy diverges since in these limits either the uncertainty in the existence of a spike in $x$ or in its timing vanishes.

\subsection{Numerical example including explicit calculation of pathwise transfer entropy}

\begin{figure*}[!htb]
\centering
\makebox[0pt]{\includegraphics[scale=0.6]{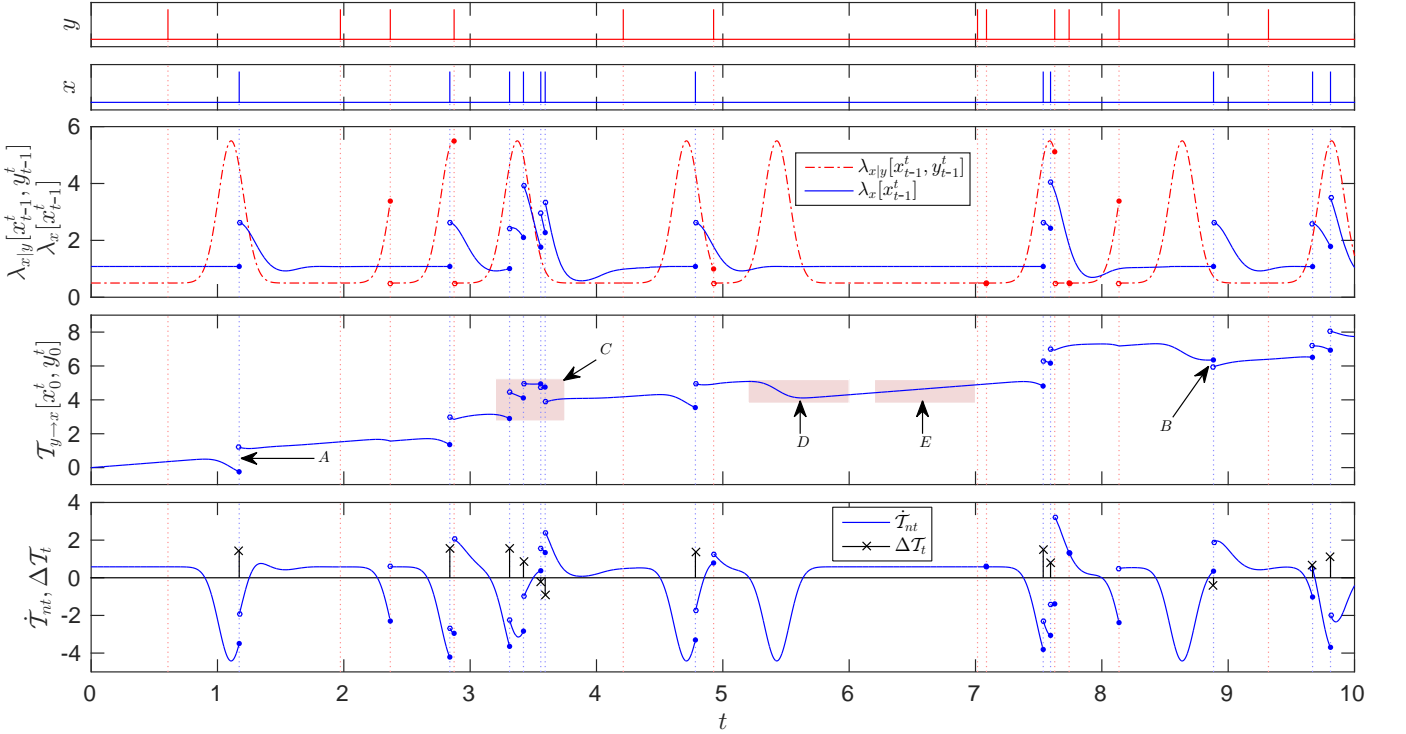}}
\caption{\label{fig2}Coupled spike trains generated using transition rates in Eq.~(\ref{rate2}) using $\lambda_x^{\text{base}}=0.5, m=5, \sigma=0.1, t_{\text{cut}}=1$ along with generated and computed values of $\lambda_{x|y}$ \& $\lambda_x$, resulting pathwise transfer entropy ($\mathcal{T}_{{y}\to{x}}$) and local contributions ($\Delta \mathcal{T}_{t}$ and $\dot{\mathcal{T}}_{nt}$). A prior history of an absence of spikes in $y$ and $x$ is assumed on the time interval $[-1,0)$. Annotations indicate pertinent features described in the text.}
\end{figure*}

In our final example, we consider a slightly more complicated process for which we compute $\lambda_x$ numerically rather than finding a limit where it can be described analytically. This allows for an illuminating graphical illustration of the pathwise transfer entropy along paths in continuous time. Once again the process is assumed to be bipartite and is defined by $\lambda_{x|y}$ and $\lambda_{y|x}$. $\lambda_x$ is then calculated numerically along spike trains (path functions) generated by the process allowing a discussion of the transfer entropy. This numerical procedure is described in Appendix \ref{appC}. The process we consider is given by the spike rates
\begin{align}
&\lambda_{y|x}[x^t_{t-s},y^t_{t-r}]=\lambda_{y}\qquad\qquad\qquad\qquad\qquad\forall\;s,r\nonumber\\
&\lambda_{x|y}[x^t_{t-s},y^t_{t-r}]=\lambda_{x|y}[y^{t}_{t-t_{\text{cut}}}]=\lambda_{x|y}(t_y^1)\quad\forall\;s,r\geq t_{\text{cut}}\nonumber\\
&\quad=\begin{cases}
\lambda_x^{\text{base}}&t_y^1\notin(0,t_{\text{cut}}]\\
\lambda_x^{\text{base}}+m\exp{[-\frac{1}{2\sigma^2}(t_y^1-\frac{t_{\text{cut}}}{2})^2]}&t_y^1\in(0,t_{\text{cut}}]\\
\qquad-m\exp{[-\frac{1}{2\sigma^2}(\frac{t_{\text{cut}}}{2})^2]}.
\label{rate2}
\end{cases}
\end{align}
where $t_y^1=t-t^y_1$ is the time \emph{since} the last spike in $y$ (where, as before, $t_1^y$ represents the time of the last spike in the relevant path history) and where, again, the system is bipartite such that both $x$ and $y$ cannot spike simultaneously. This process consists of a background rate on the target $\lambda_x^{\text{base}}$ which becomes elevated following a source spike in the regime $0\leq t_y^1< t_{\text{cut}}$. Specifically we choose this elevation to follow a Gaussian form centered on $t_y^1=t_{\text{cut}}/2$ with variance $\sigma^2$. The Gaussian is then truncated and shifted to ensure continuity in the rate function. 
One can think of this system as a hybrid Cox-renewal process. The reason being, once we consider $y$ to also be a spiking neuron (in this cases a Poisson process), $x$ can be thought of as an inhomogeneous Poisson process with rate dependent, exclusively, on the process $y$, and specifically the time since the last spike in $y$ in the manner of a renewal process.
In this example we utilize parameter $\lambda_x^{\text{base}}=0.5, m=5, \sigma=0.1, t_{\text{cut}}=1$. Two simulated spike trains along with the calculated joint \& coarse transition rates and annotated resultant pathwise transfer entropy are shown in Fig.~(\ref{fig2}). Annotations highlight important explanatory features and are commented on below.
We note that while the spiking sequences may be considered \emph{c\`adl\`ag}, the spike rates, pathwise transfer entropy, and local components are to be interpreted as \emph{c\`agl\`ad} (left continuous with right limits) since they are functionals of the right open intervals $[t_0,t)$ and $[t-1,t)$. We emphasize that $\dot{\mathcal{T}}_{nt}\neq \dot{\mathcal{T}}_{y\to x}$, the latter being undefined at target spikes and the former being the time derivative of the component which permits description in terms of local rates. We note that discontinuities in all quantities occur at spikes in either $x$ or $y$ depending on the quantity in question, but that discontinuities originating from spikes in $y$ only exist when the previous spike in $y$ is within $t_{\text{cut}}=1$ seconds of the spike in question because of the form of the rates in Eq.~(\ref{rate2}). We point out that in order to produce values for $0\leq t<1$, a prior history of an absence of spikes in $y$ and $x$ is assumed on the time interval $[-1,0)$.

A spike in $x$ during the elevated rate period where knowledge of the source process $y$ is informative in the predictability of $x$ is illustrated at point $A$ and is associated with a discontinuous increase in the pathwise transfer entropy. In contrast including knowledge of $y$ at point $B$ (outside of the elevated rate period) is misinformative and is therefore associated with a discontinuous decrease in the pathwise transfer entropy.

The cluster of target spikes at annotated region $C$ is more nuanced. At first $y$ is informative and so there is a large increase in pathwise transfer entropy with the first spike. However, the contributions associated with subsequent spikes are less significant as the rate function $\lambda_x$ begins to more accurately reflect the elevation in $\lambda_{x|y}$ due to the predicative capability it can derive from the recent spikes in its history. However, these spikes leave $\lambda_x$ elevated even once knowledge of $y$ has predicted the exit from the elevated rate period, and so $y$ is misinformative about the arrival of the final target spike in this region.

Considering instead the continuously varying non-spiking component, there are broadly two distinct situations. Decreases are associated with knowledge of the source suggesting an increase in likelihood of spikes over knowledge of just the target, but with no anticipated spike arriving. This occurs in annotated region $D$, an elevated rate period where no spikes occur. In contrast, when knowledge of the source suggests a lower likelihood of spikes over that arising from knowledge of the target alone, and no spikes occur, the inclusion of $y$ provides a better estimate leading to a positive contribution. This can occur, for this process, when there are no recent spikes in either the source or target, for example, in annotated zone E. We expect, for this finite sample, both contributions to approximately cancel because, on average over the ensemble, the non-spiking contribution must be zero as indicated by Eq.~(\ref{nonspike}). Finally we point out that because the discontinuous and continuous contributions are based on the prediction of opposite behavior (spiking vs. not spiking) whenever $\Delta \mathcal{T}_{t}(t_i)\geq 0$ then $\dot{\mathcal{T}}_{nt}(t)\leq 0$ and vice versa.

\section{Conclusion}
In this paper, we have introduced a generalization of the transfer entropy in terms of Radon-Nikodym derivatives between probability measures and an extension to continuous time systems.
For a consistent notion of transfer entropy to exist, we have emphasized that we must deal with transfer entropy rates. We have also shown, however, that the notion of a \emph{local} transfer entropy rate is not generally well defined. The natural solution, therefore, is to deal with integrated quantities which do exist. The implication is clear: transfer entropy should be understood as a \emph{dynamical} quantity accumulated along evolution of some process. Consequently, the statement ``the transfer entropy at time $t$'' is not strictly complete, but should be formally associated with a time interval over which it has been accumulated. This interpretation holds in both continuous and discrete time, where in the latter the time interval is usually ``one time step'' (for which the units are often implicit, ignored, or arbitrary). This places transfer entropy within the same family of physical quantities such as work and heat, for which there are rich accounts of their description as functionals constructed from probability measures of paths \cite{seifert_stochastic_2008,seifert_stochastic_2012,spinney_fluctuation_2013}. This underlines some of the more modern advances revealing parallels between information-theoretic and thermodynamic quantities, e.g., \cite{spin16a}.

In general, there is no obvious way to proceed with an empirical estimate of these transfer entropies for arbitrary continuous time processes aside from the brute force approach of time binning.
However, by starting from an appropriate continuous time formulation, we have pointed out that there exist classes of stochastic processes in continuous time where the constituent measures may be directly written with a finite number of variables, allowing us to sidestep time binning and its associated issues. Specifically, we have given forms for the pathwise transfer entropy and mean transfer entropy rate for arbitrary jump processes, which can be readily utilized to model spike train processes. The expression for the pathwise transfer entropy rate consists of two distinct components related to the sum of differences in local surprise associated with the transitioning behavior and an integral resulting from a continuous limit of the summation of such contributions from the non-transitioning behavior. We have also shown that the mean of the contributions arising from non-transitioning behavior must vanish such that the mean transfer entropy rate permits a simple form, promising particularly straightforward computation from empirical data. Such a result promises to be of great utility within computational neuroscience both theoretically and empirically.
In future work we will outline an estimation algorithm which can exploit the formalism presented here for spiking or point processes such that it can be applied to empirical spike (or event) timing data.
Challenges for such an algorithm center around accurately and efficiently estimating the history-dependent spike rates.
We expect such an estimator to be able to sidestep the issues associated with time binning (undersampling, etc.) specifically because our formulation permits a compressed representation of path histories in terms of spike times [i.e., $x^t_{t_0}\equiv\{t,\{t\}_0^{N_x}\}$, as per eq.~(\ref{spikepath})].

\begin{acknowledgments}
J.L. was supported through the Australian Research Council DECRA Grant No. DE160100630, and a Faculty of Engineering and IT Early Career Researcher and Newly Appointed Staff Development Scheme grant.
We thank L. Barnett and M. Wibral for helpful comments on this manuscript.
\end{acknowledgments}

\begin{appendix}
\section{Finite dimensional distributions of natural measures}
\label{appAA}
The natural measures $\mathbb{P}_{X}^{(s)}$ and $\mathbb{P}_{X|\{Y\}}^{(s,r)}$ which jointly satisfy Eqs.~(\ref{eq:eq2}), (\ref{local0}), and (\ref{avint}) possess the following structure in their finite dimensional distributions.
Given $[t^0,t^n]\subseteq\mathbb{T}\subseteq \mathbb{R}$ the (family of) finite dimensional distributions for times $(t^0<t^1<\ldots<t^n)$ is a measure on the space $((\Sigma_x)^n,\otimes^n\mathcal{X})$ and satisfies
\begin{widetext}
\begin{align}
&\mathbb{P}_{X_{0,\ldots,n}}^{\mathbb{T},(s)}\left(\bigcap_{i=0}^{n}x_{t^i}\in\mathcal{A}_i\right)=\mathbb{P}_{t^0}(x_{t^0}\in\mathcal{A}_0)\prod_{i=0}^{n-1}\mathbb{P}_{t^{i+1}}\left(x_{t^{i+1}}\in\mathcal{A}_{i+1}\biggr\rvert\bigcap_{t^j\in[t^{i}-s,t^{i}],j\geq 0}\left(x_{t^j}\in\mathcal{A}_{j}\right)\right)
\label{PX}
\end{align}
along with requisite consistency conditions thus corresponding to and implying the existence of the measure $\mathbb{P}_{X}^{\mathbb{T},(s)}$ on a suitably defined path space, $(\Omega_x^\mathbb{T},\mathcal{F}_x^\mathbb{T})$, dictating the regularity of the paths if appropriate \cite{billingsley_probability_1995}.\footnote{If, for example, $x$ has absolutely continuous sampling paths (driven perhaps by some coloured Gaussian noise) with $\Sigma_x= \mathbb{R}$, $\Omega_x^\mathbb{T}$ would be the space of continuous functions, $\mathcal{C}(\mathbb{T},\mathbb{R})$, with $\mathcal{F}_x^{\mathbb{T}}$ being the Borel sigma algebra associated with the uniform topology on $\mathcal{C}(\mathbb{T},\mathbb{R})$ such that $\mathbb{P}^{\mathbb{T},(s)}_{X}$, given the appropriate Gaussian forms for $P_{t^i}$, would be the continuous version of the extension of Eq.~(\ref{PX}).} Eq.~(\ref{PX}) should be understood as a generalization of the usual decomposition of a joint measure or density into conditional measures or densities utilized, for example, in Markov chains, but with the (not necessarily true) \emph{assumption}, $\mathbb{P}_{t'}(x_{t'}\in\mathcal{A}|\mathcal{F}_t)=\mathbb{P}_{t'}[x_{t'}\in\mathcal{A}|x^t_{t-s}]$, i.e. that given knowledge of $s$ seconds of the processes' history, further knowledge of previous history does not help in making future predictions. This amounts to a generalization of the usual Markov property $\mathbb{P}_{t'}(x_{t'}\in\mathcal{A}|\mathcal{F}_t)=\mathbb{P}_{t'}(x_{t'}\in\mathcal{A}|x_{t})$ used in the construction of more familiar entities such as the Chapman Kolmogorov equation.
By then asserting $\mathbb{T}\supseteq[t_0-\max(s,r),t)$ we define $\mathbb{P}_{X}^{(s)}[x_{t_0}^{t}\in\mathcal{A}|x_{t_0-s}^{t_0}]$ as a measure on the sub-space of functions on $[t_0,t)$, $(\Omega_x,\mathcal{F}_x)$, by appealing to regular conditional probabilities of $\mathbb{P}_{X}^{\mathbb{T},(s)}$. Similarly,
\begin{align}
&\mathbb{P}_{X_{0,\ldots,n}|\{Y\}}^{\mathbb{T},(s,r)}\left(\bigcap_{i=0}^{n}x_{t^i}\in\mathcal{A}_i\right)=\mathbb{P}_{t^0}(x_{t^0}\in\mathcal{A}_0 | y^{t^{0}}_{t^{0}-\min(t^{0}-\inf{\mathbb{T}},r)})\nonumber\\
&\qquad\qquad\qquad\qquad\times\prod_{i=0}^{n-1}\mathbb{P}_{t^{i+1}}\left[x_{t^{i+1}}\in\mathcal{A}_{i+1}\biggr\rvert \left(\bigcap_{t^j\in[t^{i}-s,t^{i}],j\geq 0}\left(x_{t^j}\in\mathcal{A}_{j}\right)\right)\cap \left(y^{t^{i}}_{t^{i}-\min(t^{i}-\inf{\mathbb{T}},r)}\right)\right]
\label{meas}
\end{align}
\end{widetext}
defines the measure $\mathbb{P}_{X|\{Y\}}^{\mathbb{T},(s,r)}$ on $(\Omega_x^\mathbb{T},\mathcal{F}_x^\mathbb{T})$, using the analogous assumption $\mathbb{P}_{t'}(x_{t'}\in\mathcal{A}|\mathcal{F}_t)=\mathbb{P}_{t'}[x_{t'}\in\mathcal{A}|x^t_{t-s},y^t_{t-r}]$. 
Again we define $\mathbb{P}_{X|\{Y\}}^{(s,r)}[x_{t_0}^{t}|x_{t_0-s}^{t_0},\{y_{t_0-r}^{t}\}]$ on $(\Omega_x,\mathcal{F}_x)$ using regular conditional probabilities of $\mathbb{P}_{X|\{Y\}}^{\mathbb{T},(s,r)}$.

\section{Derivation of pathwise transfer entropy for jump processes}
\label{appA}
Using the transition and escape rates defined in Sec. \ref{sec:jump}, we can construct path probability measures of a jump process, $x_{t_0}^t$, running from time $t'=t_0$ to time $t'=t$ that satisfy Eq.~(\ref{PX}). We reiterate that for a path that consists of $N_x$ transitions in $x$ such that transitions may be labeled by the index $i\in\{1,\ldots,N_x\}$, we write the state labels $x_i\equiv x_{t_i}$ as the state into which the system transitions at time $t_i$, with $t_0$ indicating the initial time, and $x_0=x^-_{t_0}$ the initial state.

Since we can characterize any path by an unbounded, but \emph{countable}, number of variables in this way we can directly write the probability measure for a given cylinder set. 
We do this by writing $x_{t_0}^t \in\mathcal{A}_1^{N_x}$ and understand it to mean that the path contains precisely $N_x$ transitions with $\{t_1\in\mathcal{A}_1,\ldots,t_{N_x}\in\mathcal{A}_{N_x}\}$, $\{x_1\in\mathcal{A}^1,\ldots,x_{N_x}\in\mathcal{A}^{N_x}\}$, given initial state $x_0$ at starting time $t_0$ and where $\mathcal{A}_i$ are connected subsets of $\mathbb{R}$. For simplicity we assume $(\mathcal{A}_i\cap \mathcal{A}_j)=\emptyset$, $(\mathcal{A}^i\cap \mathcal{A}^{i+1})=\emptyset$ $\forall i$ and $\inf \mathcal{A}_{i+1} > \sup \mathcal{A}_i \forall i$. 
By recognizing that we can rewrite the rates in Eqs.~(\ref{Wrates}) as $\lim_{dt\to 0}\int_{[t,t+dt]}W[{x}'|{{x}}_{t'-s}^{t'}]dt' = \lim_{dt\to 0}\mathbb{P}_{[t,t+dt]}[{x}'\in[t,t+dt]|{{x}}_{t-s}^{t}]$  we can generalize to an entire path which utilize integrals over finite time intervals by including finite probability measures of having no transition during the appropriate intervals such that
\begin{widetext}
\begin{align}
&\mathbb{P}_{X}^{(s)}[x_{t_0}^{t}\in\mathcal{A}_1^{N_x}|x^{t_0}_{t_0-s}]=\int_{\mathcal{A}_1^{N_x}}d\mathbb{P}_{X}^{(s)}[x_{t_0}^{t}|x^{t_0}_{t_0-s}]\nonumber\\
&=\sum_{x_1\in\mathcal{A}^1}\ldots \sum_{x_{N_x}\in\mathcal{A}^{N_x}}\int_{\mathcal{A}_1}dt_1\ldots\int_{\mathcal{A}_{N_x}}dt_N\left(\prod_{i=1}^{N_x}W[x_{i}|x^{t_i}_{t_i-s}]\mathbb{P}^{(s)}[x_{t'}=x_i \:\forall\: t'\in [t_{i-1},t_i)]\right) \mathbb{P}^{(s)}[x_{t'}=x_{N_x} \:\forall\: t'\in [t_{N_x},t)],
\end{align}
\end{widetext}
where the (finite) probability measures, $\mathbb{P}^{(s)}$, implicitly depend on $s$ seconds of prior history at all times. We may identify the form of $\mathbb{P}^{(s)}$ by recognizing that we must have
\begin{align}
&\frac{d\mathbb{P}^{(s)}[x_{t'}=x_j \:\forall\: t'\in [t_{j-1},t_j)]}{dt_j}\nonumber\\
&\qquad=-\lambda_x[x^{t_j}_{t_j-s}]\mathbb{P}^{(s)}[x_{t'}=x_j \:\forall\: t'\in [t_{j-1},t_j)]
\end{align}
which has solution
\begin{align}
\mathbb{P}^{(s)}[x_{t'}=x_j \:\forall\: t'\in [t_{j-1},t_j)]=\exp{\left[-\int_{t_{j-1}}^{t_j}\lambda_x[x^{t'}_{t'-s}]dt'\right]}
\end{align}
given boundary condition $\mathbb{P}^{(s)}[x_{t'}=x_j \:\forall\: t'\in [t_{j-1},t_j)]=1$ for $\lim t_j\searrow t_{j-1}$. We point out that for consistency we have $\mathbb{P}^{(s)}[x_{t'}=x_j \:\forall\: t'\in [t_{j-1},t_j)]=p_0^{(s)}[x^{t_j}_{t_{j-1}}\equiv\{t_j,\{t_{j-1},x_{t_{j-1}}\}\}|x^{t_{j-1}}_{t_{j-1}-s}]$.
Alternatively, and perhaps more intuitively, we see that such a form agrees with the limit of a time discretization where the probability of not transitioning is considered at every time step viz.,
\begin{align}
&\mathbb{P}^{(s)}[x_{t'}=x_j \:\forall\: t'\in [t_{j-1},t_j)]\nonumber\\
&\quad =\lim_{dt\to 0}\prod_{i={t_{j-1}/dt}}^{t_j/dt}(1-\lambda_x[x^{idt}_{idt-s}]dt)\nonumber\\
&\quad=\exp{\left[-\int_{t_{j-1}}^{t_j}\lambda_x[x^{t'}_{t'-s}]dt'\right]}+\mathcal{O}(dt^2)
\end{align}
where we simplify by recognizing the form of the Taylor series of the exponential to first order in $dt$.
Consequently we may write
\begin{widetext}
\begin{align}
&\mathbb{P}_{X}^{(s)}[x_{t_0}^{t}\in\mathcal{A}_1^{N_x}|x^{t_0}_{t_0-s}]=\int_{\mathcal{A}_1^{N_x}}d\mathbb{P}_{X}^{(s)}[x_{t_0}^{t}|x^{t_0}_{t_0-s}]\nonumber\\
&=\sum_{x_1\in\mathcal{A}^1}\ldots \sum_{x_{N_x}\in\mathcal{A}^{N_x}}\int_{\mathcal{A}_1}dt_1\ldots\int_{\mathcal{A}_{N_x}}dt_{N_x}\left(\prod_{i=1}^{N_x}W[x_{i}|x^{t_i}_{{t_{i}}-s}]\right)\exp{\left[-\int_{t_0}^{t}\lambda_x[x^{t'}_{t'-s}]dt'\right]}.
\label{jumpmeasure}
\end{align}
\end{widetext}
This then naturally forms a probability density for a path $x^t_{t_0}\equiv\{t,\{t,x\}_0^{N_x}\}$ with units $(\prod_{i=1}^{N_x}t_i)^{-1}$
\begin{align}
&p_{N_x}^{(s)}[x^t_{t_0}\equiv\{t,\{t,x\}_0^{N_x}\}|x^{t_0}_{t_0-s}]\nonumber\\
&\quad=\left(\prod_{i=1}^{N_x}W[x_{i}|x^{t_i}_{{t_{i}}-s}]\right)\exp{\left[-\int_{t_0}^{t}\lambda_x[x^{t'}_{t'-s}]dt'\right]}
\label{eq:probden0}
\end{align}
where again $\{t,x\}_0^{N_x}=\{\{t_0,x_0\}\dots\{t_{N_x},x_{N_x}\}\}$ meaning we can represent any path $x^t_{t_0}\equiv\{t,\{t,x\}_0^{N_x}\}$.
We note that the product term in \eq{probden0} is over the $N_x$ transitions of $x$, whilst the remaining exponentiated integral term relates to waiting times between transitions.
Expectations are taken w.r.t. this measure by performing an infinite series of integrals of the following form
\begin{widetext}
\begin{align}
\mathbb{E}_{\mathbb{P}_{X}^{(s)}}\left[f[x_{t_0}^t]\right]&=\int_{\Omega_x}f[x_{t_0}^t]d\mathbb{P}_{X}^{(s)}[x_{t_0}^t]\nonumber\\
&=f_0(x^t_{t_0}\equiv\{t,\{t_0,x_0\}\})p_0^{(s)}[x^t_{t_0}\equiv\{t,\{t_0,x_0\}\}|x^{t_0}_{t_0-s}]\nonumber\\
&+\sum_{\substack{x_1\in\Sigma_{x} \\ x_1\neq x_{0}}}\int_{t_0}^{t}dt_1f_{1}(x^t_{t_0}\equiv\{t,\{t,x\}_0^{1}\})p_{1}^{(s)}[x^t_{t_0}\equiv\{t,\{t,x\}_0^{1}\}|x^{t_0}_{t_0-s}]\nonumber\\
&+\sum_{N_x=2}^{\infty}\sum_{\substack{x_1\in\Sigma_{x} \\ x_1\neq x_{0}}}\ldots \sum_{\substack{x_{N_x}\in\Sigma_{x} \\ x_{N_x}\neq x_{N_x-1}}}\int_{t_0}^{t}dt_1\ldots\int_{t_{N_x-1}}^{t}dt_{N_x} f_{N_x}(x^t_{t_0}\equiv\{t,\{t,x\}_0^{N_x}\})p_{N_x}^{(s)}[x^t_{t_0}\equiv\{t,\{t,x\}_0^{N_x}\}|x^{t_0}_{t_0-s}]
\end{align}
where $f_i$ is the functional form that $f[x_{t_0}^t]$ takes when there are $i$ transitions in $x$ on the interval $[t_0,t)$ and where $p_i^{(s)}$ is the probability density for a path on $t'\in[t_0,t)$ that contains $i$ transitions, conditional upon the previous path function $x_{t_0-s}^{t_0}$, where transition rates utilize $s$ seconds of history dependence. Such a form is then stated more concisely through eq.~(\ref{Jsum}). We note that $p_i^{(s)}$ would also be a density with respect to $\{x_1,\ldots,x_{N_x}\}$ should $x$ be continuous. Whilst the above is formalized to include only knowledge of $x$, this can be trivially extended to include knowledge of $y$ such that we can describe the properties of $\mathbb{P}_{X|\{Y\}}^{(s,r)}$ with appropriate dependence in the transition and escape rates such that we use probability densities
\begin{align}
&p_{N_x}^{(s,r)}[x^t_{t_0}\equiv\{t,\{t,x\}_0^{N_x}\}|x^{t_0}_{t_0-s},\{y^{t}_{t_0-r}\}]=\left(\prod_{i=1}^{N_x}W[x_{i}|x^{t_i}_{{t_{i}}-s},y^{t_i}_{{t_{i}}-r}]\right)\exp{\left[-\int_{t_0}^{t}\lambda_{x|y}[x^{t'}_{t'-s},y^{t'}_{t'-r}]dt'\right]}.
\label{probden}
\end{align}\\
\\
We now have path measures which reduce to functions of the transition times which are continuous variables. The natural information theoretic interpretation leads to differential entropies, which have known issues surrounding positivity and scale invariance amongst others. However, transfer entropy, identified as a function of a Radon-Nikodym derivative, avoids these issues in all (e.g. discrete and/or continuous) potential state spaces. We form the pathwise transfer entropy by first considering the Radon-Nikoym derivative between the two measures on samples $x_{t_0}^{t}$ which must satisfy (writing $\mathbb{P}^{(s)}_X[x_{t_0}^t]$ and $\mathbb{P}^{(s,r)}_{X|\{Y\}}[x_{t_0}^t]$ as shorthand for $\mathbb{P}_{X}^{(s)}[x_{t_0}^t|x_{t_0-s}^{t_0}]$ and $\mathbb{P}_{X|\{Y\}}^{(s,r)}[x_{t_0}^t|x_{t_0-s}^{t_0},\{y^t_{t_0-r}\}]$, respectively)
\begin{align}
\mathbb{P}^{(s,r)}_{X|\{Y\}}[x_{t_0}^t\in\mathcal{A}]&=\int_{\mathcal{A}}\exp{\left[\mathcal{T}^{(s,r)}_{y\to x}[x^t_{t_0},y^t_{t_0}]\right]}d\mathbb{P}^{(s)}_X[x_{t_0}^t]=\int_{\mathcal{A}}\frac{d\mathbb{P}^{(s,r)}_{X|\{Y\}}[x_{t_0}^t]}{d\mathbb{P}^{(s)}_X[x_{t_0}^t]}d\mathbb{P}^{(s)}_X[x_{t_0}^t].
\label{eq:radnik}
\end{align}
We can compute this, heuristically, but safely in this instance, by considering the limit
\begin{align}
\frac{d\mathbb{P}_{X|\{Y\}}^{(s,r)}[x_{t_0}^t|x_{t_0-s}^{t_0},\{y^t_{t_0-r}\}]}{d\mathbb{P}_{X}^{(s)}[x_{t_0}^t|x_{t_0-s}^{t_0}]}&=\lim_{\mathcal{A}_{1}^{N_x}\to\{t,\{t,x\}_0^{N_x}\}}\frac{\mathbb{P}_{X|\{Y\}}^{(s,r)}[x_{t_0}^t\in\mathcal{A}_{1}^{N_x}|x_{t_0-s}^{t_0},\{y^t_{t_0-r}\}]}{\mathbb{P}_{X}^{(s)}[x_{t_0}^t\in\mathcal{A}_{1}^{N_x}|x_{t_0-s}^{t_0}]}\nonumber\\
&=\frac{p_{N_x}^{(s,r)}[x^t_{t_0}\equiv\{t,\{t,x\}_0^{N_x}\}|x_{t_0-s}^{t_0},\{y_{t_0-r}^{t}\}]}{p_{N_x}^{(s)}[x^t_{t_0}\equiv\{t,\{t,x\}_0^{N_x}\}|x_{t_0-s}^{t_0}]}\nonumber\\
&=\left(\prod_{i=1}^{N_x}\frac{W[x_{t_i}|x^{t_i}_{{t_{i}}-s},y^{t_i}_{{t_{i}}-r}]}{W[x_{t_i}|x^{t_i}_{{t_{i}}-s}]}\right)\exp{\left[-\int_{t_0}^{t}(\lambda_{x|y}[x^{t'}_{t'-s},y^{t'}_{t'-r}]-\lambda_x[x^{t'}_{t'-s}])dt'\right]},
\end{align}
which by comparison with Eq.~(\ref{jumpmeasure}) can be used, as expected, as a change of measure [c.f. \eq{radnik}]. The pathwise transfer entropy appearing in Eq.~(\ref{eq:local}) then directly follows.
\section{Behavior of the neuron model and calculation of its transfer entropy rate}
\label{appB}
In this appendix, we wish to give an account of the spiking neuron model in the low source spike rate, leading order in $\lambda_y$, regime. To this end, we present both the transfer entropy rate to leading order in $\lambda_y$ and a scheme for approximating the pathwise transfer entropy, again to first order in $\lambda_y$. The model set up specifies a constant $\lambda_{y|x}=\lambda_y$ outside of the refractory period of length $\tau^r$ in $y$, and specifies that $\lambda_{x|y}=-\tau^{-1}\ln[1-a]$ up until the first spike in $x$ up to $\tau$ seconds after a spike in $y$ and zero at all other times or when $x$ is within its refractory period also of length $\tau^r$. While these aspects are immediately defined by the model, $\lambda_x$ is not and so we must calculate its value, up to $\mathcal{O}(\lambda_y)$, for our purposes.\\
\\
To do so we formulate the spike rate in $x$, informally, but safely, by the expression
\begin{align}
&\lambda_x[x^{t}_{t-q}\equiv\{q,t,\{t\}_1^{N_x}\}]=\lim_{\substack{\mathcal{A}_1^{N_x}\to\{q,t,\{t\}_1^{N_x}\}\\dt\to 0}}\frac{(dt)^{-1}\mathbb{P}_{X}[x_{\text{spike}}\in[t,t+dt]\cap x_{t-q}^t\in \mathcal{A}_1^{N_x}] }{\mathbb{P}_{X}[x_{t-q}^t\in\mathcal{A}_1^{N_x}]}
\label{main}
\end{align}
where $\{q>0\}\in\mathbb{R}$. We may represent the whole joint path by $\{x^t_{t-q},y^t_{t-q}\}\equiv\{q,t,\{t^x\}_1^{N_x},\{t^y\}_1^{N_y}\}$ and thus represent the denominator, dropping the explicit equivalence in earlier notation, as
\begin{align}
\mathbb{P}_{X}[x_{t-q}^t\in\mathcal{A}_1^{N_x}]&=\int_{\mathcal{A}_1^{N_x}\times{\Omega_y^{[t-q,t]}}}d\mathbb{P}_{XY}[x_{t-q}^t,y_{t-q}^t]\nonumber\\
&=\int_{\mathcal{A}_1}dt^{x}_1\ldots \int_{\mathcal{A}_{N_x}}dt^{x}_{N_x}p_{N_x,0}(q,t,\{t^x\}_1^{N_x})+\int_{t-q}^{t}dt^y_1\int_{\mathcal{A}_1}dt^{x}_1\ldots \int_{\mathcal{A}_{N_x}}dt^{x}_{N_x}p_{N_x,1}(q,t,\{t^x\}_1^{N_x},t^y_1)\nonumber\\
&+\sum_{N_y=2}^{\infty}\int_{t-q}^{t}dt^y_1\ldots\int_{t_{N_y-1}}^{t}dt^y_{N_y}\int_{\mathcal{A}_1}dt^{x}_1\ldots \int_{\mathcal{A}_{N_x}}dt^{x}_{N_x} p_{N_x,N_y}(q,t,\{t^x\}_1^{N_x},\{t^y\}_1^{N_y})
\label{int}
\end{align}
where $p_{i,j}$ is the probability density function for a path with $i$ spikes in $x$ and $j$ spikes in $y$ and $\Omega_y^{[t-q,t]}$ is the entire relevant path space for trajectories in $y$ on $[t-q,t]$. The numerator is then given by

\begin{align}
\lim_{dt\to 0}&\frac{1}{dt}\mathbb{P}_{X}[x_{\text{spike}}\in [t,t+dt]\cap x_{t-q}^t\in\mathcal{A}_1^{N_x}]\nonumber\\
&=\lim_{dt\to 0}\frac{1}{dt}\int_{\mathcal{A}_1^{N_x}\times{\Omega_y^{[t-q,t+dt]}}}d\mathbb{P}_{XY}[x_{t-q}^{t+dt},y^{t+dt}_{t-q}|t^x_{N_x+1}\in [t+dt]]\nonumber\\
&=\int_{\mathcal{A}_1}dt^{x}_1\ldots \int_{\mathcal{A}_{N_x}}dt^{x}_{N_x}\lambda_{x|y}^{N_x,0}(q,t,\{t^x\}_1^{N_x})p_{N_x,0}(q,t,\{t^x\}_1^{N_x})\nonumber\\
&+\int_{t-q}^{t}dt^y_1\int_{\mathcal{A}_1}dt^{x}_1\ldots \int_{\mathcal{A}_{N_x}}dt^{x}_{N_x}\lambda_{x|y}^{N_x,1}(q,t,\{t^x\}_1^{N_x},t^y_1)p_{N_x,1}(q,t,\{t^x\}_1^{N_x},t^y_1)\nonumber\\
&+\sum_{N_y=2}^{\infty}\int_{t-q}^{t}dt^y_1\ldots\int_{t^y_{N_y-1}}^{t}dt^y_{N_y}\int_{\mathcal{A}_1}dt^{x}_1\ldots \int_{\mathcal{A}_{N_x}}dt^{x}_{N_x}\lambda_{x|y}^{N_x,N_y}(q,t,\{t^x\}_1^{N_x},\{t^y\}_1^{N_y})p_{N_x,N_y}(q,t,\{t^x\}_1^{N_x},\{t^y\}_1^{N_y})
\label{num_appc}
\end{align} 
\end{widetext}
where $\lambda_{x|y}^{i,j}$ is the spike rate in $x$ given $i$ spikes in the history of $x$ and $j$ spikes in the history of $y$ and $\Omega_y^{[t-q,t+dt]}$ is the entire relevant path space for trajectories in $y$ on $[t-q,t+dt]$. Since, in our example, $p_{i,N_y}\propto \lambda_y^{N_y}$, the integrated terms in eqs.~(\ref{int}) and (\ref{num_appc}) containing $p_{i,N_y}$ must also be $\mathcal{O}(\lambda_y)$ (with higher order corrections with origin in the refractory periods). Consequently, when estimating $\lambda_x$, up to $\mathcal{O}(\lambda_y)$, we can truncate these infinite series. Where we are permitted to truncate the series is then determined by the chosen history dependence of $\lambda_x[x^{t}_{t-q}]$. In our example each spike in $x$ must be preceded by a spike in $y$. Consequently, since we are considering the $\mathcal{O}(\lambda_y)$ regime,  such that we need only $\mathcal{O}(\lambda_y)$ contributions, we consider only the dominant path histories where the limit is valid. As such we neglect the path histories with probability densities $\propto \lambda_y^2$ and higher ($N_x\geq 2$) and consider only (up to) one spike in the history of $x$, $N_x\leq 1$, such that we may consider $\lambda_x[x^{t}_{t-q}\equiv \{q,t,t^x_1\}]=\lambda_x^1(q,t,t^x_1)$ or $\lambda_x[x^{t}_{t-q}\equiv\{q,t\}]=\lambda_x^0(q,t)$ (where $\lambda_x^i$ is the spike rate given $i$ spikes in the history of $x$). \\

We note that, despite this restriction, since $\lambda_x$ effectively provides a weighted estimate of being within the $\tau$ second long elevated rate period that follows a spike in $y$, knowledge of additional historical spikes in $x$ sufficiently far in the past cannot have an effect on the coarse grained rate $\lambda_x$. This is because any inference from these distant spikes cannot change the likelihood of currently being in an elevated rate period. By considering the most recent time a hidden spike in $y$ can be associated with an additional historical spike in $x$ and its potential impact on subsequent spiking rates we understand that any previous spikes in $x$, $\tau^r+\tau$ seconds or more prior to the time of the earliest spike in $x$ in the explicitly considered history, cannot effect its functional form. This is because this is the latest time the additional previous spike in $x$ can occur, for which the most recent possible associated causative spike in $y$ can have occurred, which gives $y$ time to subsequently pass through its refractory period and then spike again, causing the subsequent spike in the history of $x$, without the uncertainty in its timing being reduced below the default $\tau$ seconds length of the elevated rate window because of that previous refractory period. Since in the low source spike rate [$\mathcal{O}(\lambda_y)$] regime, spikes in $y$, and thus $x$, are increasingly uncommon, cases where the time between any previous spike and the one included in its history are less than $\tau^r+\tau$ are suitably rare so long as $\tau^r+\tau\ll\lambda_y^{-1}$.  \\
\\
We proceed with the $N_x=1$ case, returning later to the simpler $N_x=0$ case. To consider the spike rate with a history $N_x=1$ we must include the spike in $y$ that preceded the spike in the history of $x$, but also the spike that precedes the (potential) spike in question at time $t$ meaning all terms below $N_y=2$ in the numerator and the $N_y=0$ term in the denominator must vanish. More concisely we recognize that $\lambda_{x|y}^{i\geq j,j}=0$ and $p_{i> j,j}=0$ in the infinite series. Expanding the surviving terms in the series about $\lambda_y=0$, understanding that the integrals over terms in $p_{i,j}$ are to leading order $\propto\lambda_y^{j}$, we recognize that all first order terms in $\lambda_y$ for $\lambda_x^1$ are contained in the expression 
\begin{align}
 &\lambda^1_x(q,t,t^x_1)\nonumber\\
 &=\frac{\int_{t-q}^t dt^y_1\int_{t^y_1}^t dt^y_2 \lambda_{x|y}^{1,2}(q,t,t^x_1,\{t^y\}_1^2)p_{1,2}(q,t,t^x_1,\{t^y\}_1^2)}{\int_{t-q}^t dt^y_1p_{1,1}(q,t,t^x_1,t^y_1)}.
 \end{align}
 In our example, we recognize that a spike in $x$ must occur within $\tau$ seconds of a spike in $y$ so we may rewrite this by fixing the timing of the (previous) single spike in $x$, $t^x_1=0$ and set $q=t+\tau$ such that we have a result valid for all $q>t+\tau$, leaving $t$ as the only free variable. Consequently we drop the dependence on $q=t+\tau$ and write
 \begin{align}
 &\lambda_x^1(t,t^x_1=0) \nonumber\\
 &=\frac{\int_{-\tau}^t dt^y_1\int_{t^y_1}^t dt^y_2 \lambda_{x|y}^{1,2}(t,t^x_1=0,\{t^y\}_1^2)p_{1,2}(t,t^x_1=0,\{t^y\}_1^2)}{\int_{-\tau}^t dt^y_1p_{1,1}(t,t^x_1=0,t^y_1)}.
 \end{align}
Finally, we point out that the same regime where spikes occurring less than $\tau^r+\tau$ seconds before $t^x_1=0$ are negligibly rare allows us to assert that in the probability densities above, both $x$ and $y$ are not in their refractory periods at $t=-\tau$. Consequently, because $y$ is otherwise independent of $x$, we can directly write the above probability densities on the interval $[-\tau,t)$, which are given by
 \begin{align}
 &p_{1,1}(t,t^x_1=0,t^y_1)\nonumber\\
 &=\begin{cases}
 \lambda_y \lambda_{x|y}^e e^{-\lambda_y(t^y_1+\tau)}&t^y_1\in[-\tau,0)\\
 \quad\times e^{-\lambda_y(t-\text{Min}[t,t^y_1+\tau^r])}e^{\lambda_{x|y}^et^y_1},&\\
 0,&t^y_1\notin[-\tau,0).
 \end{cases}
 \end{align}
 \begin{align}
 &\lambda_{x|y}^{1,2}(t,t^x_1=0,\{t^y\}_1^2)p_{1,2}(t,t^x_1=0,\{t^y\}_1^2)\nonumber\\&=\begin{cases}
 (\lambda_y \lambda_{x|y}^e)^2 e^{-\lambda_y(t^y_1+\tau)}&t^y_1\in[-\tau,0) \\
\times e^{-\lambda_y(t^y_2-(t^y_1+\tau^r))}&t^y_2\in[\text{Max}[\tau^r,t-\tau],t)\\
\times e^{-\lambda_y(t-\text{Min}[t,t^y_2+\tau^r])}&\\
 \times e^{\lambda_{x|y}^e t^y_1}e^{-\lambda_{x|y}^e\text{Max}[\tau,t-t^y_2]},&\\
 0,&t^y_1\notin[-\tau,0)\\
 &\text{or}\\
 &t^y_2\notin[\text{Max}[\tau^r,t-\tau],t).
 \end{cases}.
 \end{align}
 We can implement such forms via the integrals
 \begin{align}
 &\int_{-\tau}^t dt^y_1p(t,t^x_1=0,t^y_1)\nonumber\\
 &=\int_{-\tau}^{0}dt^y_1 \lambda_y \lambda_{x|y}^ee^{-\lambda_y(t^y_1+\tau)}e^{-\lambda_y(t-\text{Min}[t,t^y_1+\tau^r])}e^{\lambda_{x|y}^et^y_1}\\
 &\int_{-\tau}^t dt^y_1\int_{t^y_1}^t dt^y_2 \lambda_{x|y}^{1,2}(t,t^x_1=0,\{t^y\}_1^2)p(t,t^x_1=0,\{t^y\}_1^2)\nonumber\\
 &=\int_{-\tau}^{0}dt^y_1\int_{\text{Min}[t,\text{Max}[\tau^r,t-\tau]]}^{t}dt^y_2 (\lambda_y \lambda_{x|y}^e)^2 e^{-\lambda_y(t^y_1+\tau)}\nonumber\\
 &\qquad\qquad\times e^{-\lambda_y(t^y_2-(t^y_1+\tau^r))}e^{-\lambda_y(t-\text{Min}[t,t^y_2+\tau^r])}\nonumber\\
 &\qquad\qquad\times e^{\lambda_{x|y}^et^y_1}e^{-\lambda_{x|y}^e\text{Max}[\tau,t-t^y_2]}\nonumber\\
 &=\int_{\text{Min}[t,\text{Max}[\tau^r,t-\tau]]}^{t}dt^y_2 \int_{-\tau}^{0}dt^y_1(\lambda_y \lambda_{x|y}^e)^2 e^{-\lambda_y(t^y_2+(\tau-\tau^r))}\nonumber\\
 &\qquad\qquad\qquad\times e^{-\lambda_y(t-\text{Min}[t,t^y_2+\tau^r])}e^{\lambda_{x|y}^et^y_1}e^{-\lambda_{x|y}^e(t-t^y_2)}\nonumber\\
 &=\lambda_y^2 \lambda_{x|y}^e \left(e^{\lambda_{x|y}^e\tau}-1\right)e^{-\lambda_{x|y}^e(t+\tau)-\lambda_y(t+\tau-\tau^r)} \nonumber\\
 &\qquad\times \int_{\text{Min}[t,\text{Max}[\tau^r,t-\tau]]}^{t}dt^y_2e^{\lambda_{x|y}^et^y_2-\lambda_y(t^y_2-\text{Min}[t,t^y_2+\tau^r])}.
 \end{align}
 We note the assumption of $\tau^r\geq \tau$ allows us to swap the order of the integrals as the second spike in $y$ is rendered independent of $t^y_1$. Now, from the refractory period constraint we know that $\lambda_x(t)=0$ for $t<\tau^r$ so we can ignore the computation for such a regime. Consequently, for the denominator, we can write
  \begin{align}
 &\int_{-\tau}^{0}dt^y_1 \lambda_y \lambda_{x|y}^e e^{-\lambda_y(t^y_1+\tau)}e^{-\lambda_y(t-\text{Min}[t,t^y_1+\tau^r])}e^{\lambda_{x|y}^et^y_1}\nonumber\\
 &=\int_{-\tau}^{0}dt^y_1 \lambda_y \lambda_{x|y}^e e^{-\lambda_y(t^y_1+\tau)}e^{-\lambda_y(t-t^y_1+\tau^r)}e^{\lambda_{x|y}^et^y_1},\; t\geq \tau^r\nonumber\\
&= \lambda_ye^{-\lambda_{x|y}^e\tau-\lambda_y(t+\tau-\tau^r)}(e^{\lambda_{x|y}^e\tau}-1),\quad t\geq \tau^r
 \label{eq:appADenominator}
 \end{align}
   The numerator, however, is more complicated and can be written
   \begin{align}
   &\lambda_y^2 \lambda_{x|y}^e (e^{\lambda_{x|y}^e\tau}-1)e^{-\lambda_{x|y}^e(t+\tau)-\lambda_y(t+\tau-\tau^r)}\nonumber\\
   &\times\int_{\text{Min}[t,\text{Max}[\tau^r,t-\tau]]}^{t}dt^y_2e^{\lambda_{x|y}^et^y_2-\lambda_y(t^y_2-\text{Min}[t,t^y_2+\tau^r])}\nonumber\\
 &=\lambda_y^2 \lambda_{x|y}^e (e^{\lambda_{x|y}^e\tau}-1)e^{-\lambda_{x|y}^e(t+\tau)-\lambda_y(t+\tau-\tau^r)}\nonumber\\
 &\times
  \begin{cases}
 0,&t<\tau^r\\\\
 \int_{\tau^r}^te^{\lambda_{x|y}^et^y_2-\lambda_y(t^y_2-t)}, &\tau^r\leq t<\tau^r+\tau\\\\
 \int_{t-\tau}^te^{\lambda_{x|y}^et^y_2-\lambda_y(t^y_2-t)}, &t\geq\tau^r+\tau
 \end{cases}\nonumber\\
  &=\begin{cases}
 0,&t<\tau^r\\\\
 \lambda_y^2\lambda_{x|y}^e(\lambda_{x|y}^e-\lambda_y)^{-1}, &\tau^r\leq t<\tau^r+\tau\\
 \times (e^{\lambda_{x|y}^e\tau}-1)e^{-(\lambda_{x|y}^e+\lambda_y)(t+\tau)}\\
 \times (e^{\lambda_{x|y}^et+\lambda_y\tau^r}-e^{\lambda_{x|y}^e\tau^r+\lambda_yt})\\\\
 \lambda_y^2\lambda_{x|y}^e(\lambda_{x|y}^e-\lambda_y)^{-1}, &t\geq\tau^r+\tau\\
 \times (e^{\lambda_{x|y}^e\tau}-1)(e^{\lambda_{x|y}^e\tau}-e^{\lambda_y\tau})\\
 \times e^{-2\lambda_{x|y}^e\tau-\lambda_y(t+\tau-\tau^r)}
 \end{cases}.
 \label{eq:appANumerator}
 \end{align}
 Considering the ratio of the results \eq{appANumerator} and \eq{appADenominator}, inserting $\lambda_{x|y}^e=-\tau^{-1}\ln[1-a]$ and discarding all $\mathcal{O}(\lambda_y^2)$ terms and higher we find
 \begin{align}
 &\lambda_x^1(t,t^x_1=0)=\nonumber\\
 &\begin{cases}
 0,&0\leq t<\tau^r\\
 \left(1-(1-a)^{\frac{t-\tau^r}{\tau}}\right)\lambda_y,&\tau^r\leq t<\tau^r+\tau\\
 a\lambda_y,&t\geq \tau^r+\tau
 \end{cases}\nonumber\\
 &+\mathcal{O}(\lambda_y^2).
 \end{align}
 Returning to the $N_x=0$ case we can avoid a similar, albeit simpler, calculation by recognizing that we must have, by continuity arguments, $\lambda_x^0(t)=a\lambda_y$. This then allows one to calculate the pathwise transfer entropy contributions set out in \eq{local} up to $\mathcal{O}(\lambda_y)$ wherever the interspike intervals are greater than $\tau^r+\tau$. \\
 \\
 To compute the transfer entropy rate we have, equivalently,
 \begin{align}
 \dot{T}_{y\to x}&=\frac{1}{(t-t_0)}\mathbb{E}_{P}\left[ \mathcal{T}_{y\to x}[x_{t_0}^t,y_{t_0}^t]\right]\nonumber\\
 &=\frac{1}{(t-t_0)}\mathbb{E}_{P}\left[\sum_{i=1}^{N_x}\Delta \mathcal{T}_{t}(t_i)\right]\nonumber\\
 &=\frac{1}{(t-t_0)}\mathbb{E}_{P}\left[\sum_{i=1}^{N_x}\ln{\frac{\lambda_{x|y}^e}{\lambda_{x}[x_{t_i-(\tau+\tau^r)}^{t_i}]}}\right]\\
 \dot{T}_{y\to x}&=\mathbb{E}_{P}\left[ \Delta\mathcal{T}_{t}(t)\right]\nonumber\\
  &=\mathbb{E}_{P}\left[(1-\delta_{x_t^-x_t})\ln{\frac{\lambda_{x|y}^e}{\lambda_{x}[x_{t-(\tau+\tau^r)}^{t}]}}\right].
  \label{avdef2}
 \end{align}
 Considering only $\mathcal{O}(\lambda_y)$ contributions allows for up to one transition in $y$ in the path measure such that the only significant terms comprising the integral of the form in Eq.~(\ref{int}) are $p_{0,1}(t,t^y_1)$ and $p_{1,1}(t,t^x_1,t^y_1)$ (since every spike in $x$ must be preceded by one in $y$), but with only the latter leading to transitions in $x$ and thus any transition contributions $\Delta\mathcal{T}_{t}$. Taking the definition of the transfer entropy rate in Eq.~(\ref{avdef2}) we may consequently write
 \begin{widetext}
 \begin{align}
 \dot{T}_{y\to x}(t)&=\lim_{t_0\to-\infty}\int_{t_0}^{t}dt^y_1p_{0,1}(t,t^y_1)\cdot 0+\lim_{t_0\to-\infty}\int_{t_0}^{t}dt^x_1\int_{t_0}^{t}dt^y_1p_{1,1}(t,t^x_1,t^y_1)\delta(t^x_1-t)\ln{\left[-\frac{\ln{[1-a]}}{\lambda_x^0(t)\tau}\right]}\nonumber\\
 &=\lim_{t_0\to-\infty}\int_{t_0}^{t}dt^y_1p_{1,1}(t,t^x_1=t,t^y_1)\ln{\left[-\frac{\ln{[1-a]}}{\lambda_x^0(t)\tau}\right]}=\int_{t-\tau}^{t}dt^y_1\lambda_{x|y}^e\lambda_ye^{-\lambda_y(t^y_1-(t-\tau))-\lambda_{x|y}^e(t-t^y_1)}\ln{\left[-\frac{\ln{[1-a]}}{\lambda_x^0(t)\tau}\right]}\nonumber\\
 &=(1-e^{-\lambda_{x|y}^e\tau})\lambda_y\ln{\left[-\frac{\ln{[1-a]}}{\lambda_x^0(t)\tau}\right]}+\mathcal{O}(\lambda_y^2)=a\lambda_y\ln{\left[-\frac{\ln{[1-a]}}{a\lambda_y\tau}\right]}+\mathcal{O}(\lambda_y^2).
\end{align}
For completeness we may equivalently write the former definition, acknowledging that in the $\mathcal{O}(\lambda_y)$ regime we have $N_x\leq 1$,
\begin{align}    
 \dot{T}_{y\to x}&=\frac{1}{(t-t_0)}\int_{t_0}^{t}dt^x_1\int_{t_0}^{t}dt^y_1p_{1,1}(t,t^x_1,t^y_1)\ln{\left[-\frac{\ln{[1-a]}}{\lambda_x^0(t^x_1)\tau}\right]}+\mathcal{O}(\lambda_y^2)\nonumber\\
 &=\frac{1}{(t-t_0)}\int_{t_0}^{t}dt^x_1p_1(t,t^x_1)\ln{\left[-\frac{\ln{[1-a]}}{\lambda_x^0(t^x_1)\tau}\right]}+\mathcal{O}(\lambda_y^2).
 \end{align}
 \end{widetext}
 We can write the probability density
 \begin{align}
 p_1(t,t')&=e^{-\int_{t_0}^{t'}\lambda_x^0(t'')dt''}\lambda_x^0(t')e^{-\int_{t'}^t\lambda_x^1(t'',t')dt''}\nonumber\\
 &=\lambda^0_x(t')+\mathcal{O}(\lambda_y^2).
 \end{align}
Once again, the continuity requirements dictate that $\lambda_x^0(t')=a\lambda_y=\lambda_x^1(t'',t'),\:t'-t''>\tau+\tau^r$ so that
 \begin{align}
 \dot{T}_{y\to x}=&\frac{1}{(t-t_0)}\int_{t_0}^{t}dt^x_1\lambda^0_x(t^x_1)\ln{\left[-\frac{\ln{[1-a]}}{\lambda^0_x(t^x_1)\tau}\right]}+\mathcal{O}(\lambda_y^2)\nonumber\\
    =&a\lambda_y\ln{\left[-\frac{\ln{[1-a]}}{a\lambda_y\tau}\right]}+ \mathcal{O}(\lambda_y^2).
 \end{align}

\section{Numerical scheme for arbitrary spiking process}
\label{appC}
Here, we present a numerical scheme for computing the coarse grained spike rate given a bipartite co-spiking system. We imagine that in such systems the behavior, of the joint system, at time $t$, is completely described by the conditional spike rates $\lambda_{x|y}[x^t_{t-\tau^{xx}},y^t_{t-\tau^{xy}}]$ and $\lambda_{y|x}[x^t_{t-\tau^{yx}},y^t_{t-\tau^{yy}}]$ such that the parameters $\tau^{xx}$, $\tau^{xy}$, $\tau^{yx}$ and $\tau^{yy}$ represent a finite reliance on the past in a manner analogous to a Markov order in discrete time systems. A true Markov system is achieved in the limit of these quantities going to zero. However, when calculating the spike rate $\lambda_x$ without knowledge of $y$, the spike rates may have, in principle, an infinite dependence on its past owing to the correlations that arise from the bi-directional influence between the two. Taking our previously established definition of the coarse grained spike rate in the form of Eq.~(\ref{main}) in $x$, $\lambda_x[x^t_{t-s}\equiv\{s,t,\{t^x\}_1^{N_x}\}]$, we introduce, for brevity, the notation
\begin{align}
&\int dy^t_{t_0}f(y^t_{t_0})=f_0(y^t_{t_0}\equiv\{t_0,t\})\nonumber\\
&+\int_{t_0}^tdt^y_1f_{1}(y^t_{t_0}\equiv\{t_0,t,t^y_1\})\nonumber\\
&+\sum_{N_y=2}^{\infty}\int_{t_0}^tdt^y_1\ldots\int_{t^y_{N_y-1}}^tdt^y_{N_y}f_{N_y}(y^t_{t_0}\equiv\{t_0,t,\{t^y\}_1^{N_y}\})
\label{infint}
\end{align} 
such that
\begin{align}
&\lambda_x[x^t_{t-s}\equiv\{s,t,\{t^x\}_1^{N_x}\}]\nonumber\\
&=\frac{\int dy^t_{t-s} \lambda_{x|y}^{N_x,\cdot}(\{t^x\}_1^{N_x},y^t_{t-s})p_{N_x,\cdot}(\{t^x\}_1^{N_x},y^t_{t-s})}{\int dy^t_{t-s}p_{N_x,\cdot}(\{t^x\}_1^{N_x},y^t_{t-s})}
\end{align}
such that $p_{N_x,\cdot}$ represents the probability densities (and analogously $\lambda_{x|y}^{N_x,\cdot}$ for spike rates) used in the implicit sum over $p_{N_x,j}$ for paths that contain $N_x$ spikes in $x$ and $j$ spikes in $y$ over a process of $s$ seconds duration. However, given that we can only construct probability densities from conditional spike rates we must, in general, always specify the relevant conditioning, i.e., we cannot write $p_{N_x,j}(\{t^x\}_1^{N_x},\{t^y\}_1^{j})$ but instead must write, by virtue of the process being bipartite, $p_{N_x,j}(\{t^x\}_1^{N_x},\{t^y\}_1^{j}|x^{t-s}_{t-(s+A)},y^{t-s}_{t-(s+B)})$ where $A=\max(\tau^{xx},\tau^{xy}), B=\max(\tau^{yy},\tau^{yx})$. Using such densities, and integrating over all $\{t^y\}_1^{j}$ would unavoidably lead to dependence in the calculated spike rate on $x^{t-s}_{t-(s+A)},y^{t-s}_{t-(s+B)}$ which cannot, generally, be guaranteed not to change its value. Instead, we must recognize that we cannot remove conditioning on some previous spike history, since to integrate over it introduces more conditional spike history, and instead must render it irrelevant to our calculation. To do so we recognize that because we have specified strict Markov orders in $\lambda_{x|y}$ and $\lambda_{y|x}$, any additional dependence in the coarse grained spike rate arises from correlation with the past and thus must decay with that correlation. Consequently, we write
\begin{equation}
\lambda_x[x^t_{t-s}]=\lim_{s'\to\infty}\lambda_{x|y}[x^t_{t-s},x^{t-s'}_{t-(s'+A)},y^{t-s'}_{t-(s'+B)}]
\end{equation}
which can be achieved, approximately with finite $s'$, by integrating over all sequences for $y^{t}_{t-s'}$ and $x^{t-s}_{t-s'}$ using the probability densities $p_{N_x,\cdot,\cdot}[\{t^x\}_{N'_x+1}^{N'_x+N_x},x^{t-s}_{t-s'},y^t_{t-s'}|x^{t-s'}_{t-(s'+A)},y^{t-s'}_{t-(s'+B)}]$ indicating the set of probability densities of the form $p_{N_x,N'_x,N_y}[x^{t}_{t-s}\equiv\{s,t,\{t^x\}_{N'_x+1}^{N'_x+N_x}\},x^{t-s}_{t-s'}\equiv\{s',s,\{t^x\}_{1}^{N'_x}\},y^t_{t-s'}\equiv\{s',t,\{t^y\}_1^{N_y}\}|x^{t-s'}_{t-(s'+A)},y^{t-s'}_{t-(s'+B)}]$. As such we may utilize the following representation for $\lambda_x[x^{t}_{t-s}]$
\begin{widetext}
\begin{align}
&\lambda_x(x^t_{t-s}\equiv\{s,t,\{t^x\}_1^{N_x}\})\nonumber\\
&=\lim_{s'\to\infty}\frac{\int dy^t_{t-s'}\int dx^{t-s}_{t-s'} \lambda_{x|y}^{N_x,\cdot,\cdot}[\{t^x\}_1^{N_x},x^{t-s}_{t-s'},y^{t}_{t-s'}]p_{N_x,\cdot,\cdot}[\{t^x\}_1^{N_x},x^{t-s}_{t-s'},y^{t}_{t-s'}|x^{t-s'}_{t-(s'+A)},y^{t-s'}_{t-(s'+B)}]}{\int dy^t_{t-s'}\int dx^{t-s}_{t-s'}p_{N_x,\cdot,\cdot}[\{t^x\}_1^{N_x},x^{t-s}_{t-s'},y^{t}_{t-s'}|x^{t-s'}_{t-(s'+A)},y^{t-s'}_{t-(s'+B)}]}
\end{align}
where
\begin{align}
&p_{N_x,N'_x,N_y}[x^{t}_{t-s}\equiv\{s,t,\{t^x\}_{N'_x+1}^{N'_x+N_x}\},x^{t-s}_{t-s'}\equiv\{s',s,\{t^x\}_{1}^{N'_x}\},y^t_{t-s'}\equiv\{s',t,\{t^y\}_1^{N_y}\}|x^{t-s'}_{t-(s'+A)},y^{t-s'}_{t-(s'+B)}]\nonumber\\
&=p_{N_x+N'_x,N_y}[x^t_{t-s'}\equiv\{s',t,\{t^x\}_{1}^{N'_x+N_x},\{t^y\}_1^{N_y}\}\}|x^{t-s'}_{t-(s'+A)},y^{t-s'}_{t-(s'+B)}]\nonumber\\
&=p_{N_x+N'_x}^{(\tau^{xx},\tau^{xy})}[x^t_{t-s'}\equiv\{s',t,\{t^x\}_{1}^{N'_x+N_x}\}|x^{t-s'}_{t-(s'+\tau^{xx})},\{y^t_{t-s'}\equiv\{s',t,\{t^y\}_1^{N_y}\},y^{t-s'}_{t-(s'+\tau^{xy})}\}]\nonumber\\
&\quad\times p_{N_y}^{(\tau^{yx},\tau^{yy})}[y^t_{t-s'}\equiv\{s',t,\{t^y\}_{1}^{N_y}\}|y^{t-s'}_{t-(s'+\tau^{yy})},\{x^t_{t-s'}\equiv\{s',t,\{t^x\}_1^{N_x+N'_x}\},x^{t-s'}_{t-(s'+\tau^{yx})}\}]
\end{align}
\end{widetext}
from the bipartite property of the process with the last line expressible by two probability densities of the form in Eq.~(\ref{probden}). This is then a series of (nested) summations and integrals which can be readily approximated using a discrete time scheme. Naturally, if capturing all possible path dependence in $x$, such that $s=s'\to\infty$, the path integral over $x$ is omitted.\\
\\
Discussing the practicalities of implementing such a process becomes cumbersome in the general case so we reduce the problem to the special case used in the numerical spiking example, but note that the technique would be analogous. In the example, the target $x$ depends only on the history of the source $y$, the source process $y$ is independent of the target process $x$, the source is Markov, and because the process can only ever spike from the unspiked state, the Markovian property is equivalent to complete independence of its history. This has the consequence that we may consider $\tau^{yy}\searrow 0$, $\tau^{yx} \searrow 0$ and $\tau^{xx}\searrow 0$ hereafter denoted $0^{+}$ (we also note that in our specific example we have $\tau^{xy}=t_{\rm cut}$). This also lets us fully specify all quantities involved in the construction of $p_{N_x,N_y}(x^t_{t-\tau^{xy}}\equiv\{\tau^{xy},t,\{t^x\}_1^{N_x}\},y^{t}_{t-2\tau^{xy}}\equiv\{\tau^{xy},t,\{t^y\}_1^{N_y}\})$ without conditioning such that we can write
\begin{widetext}
\begin{align}
&p_{N_x,N_y}(x^t_{t-\tau^{xy}}\equiv\{\tau^{xy},t,\{t^x\}_1^{N_x}\},y^{t}_{t-2\tau^{xy}}\equiv\{\tau^{xy},t,\{t^y\}_1^{N_y}\})\nonumber\\
&=p_{N_y}^{(0^{+},0^{+})}(y^{t}_{t-2\tau^{xy}}\equiv\{\tau^{xy},t,\{t^x\}_1^{N_y}\})p_{N_x}^{(0^{+},\tau^{xy})}(x^t_{t-\tau^{xy}}\equiv\{\tau^{xy},t,\{t^x\}_1^{N_x}\}|\{y^{t}_{t-2\tau^{xy}}\equiv\{\tau^{xy},t,\{t^x\}_1^{N_x}\}\}).
\label{biprep}
\end{align}
\end{widetext}
Notably, the independence of $y$ from $x$ provides conditions where the conditional probability density defined in the manner of Eq.~(\ref{meas}) aligns with the conditional probability density in the usual sense. Next we recognize that the independence of $x$ from its history and $y$ from $x$ means there is no mechanism for feedback from $x$ to itself, meaning that we have the property
\begin{align}
&\lambda_x[x^t_{t-s}]=\lambda_x[x^t_{t-\tau^{xy}}]\qquad \forall\; s\geq\tau^{xy}\nonumber\\
&=\frac{\int dy^t_{t-2\tau^{xy}} \lambda_{x|y}^{N_x,\cdot}(y^t_{t-\tau^{xy}})p_{N_x,\cdot}[\{t^x\}_1^{N_x},y^t_{t-2\tau^{xy}}]}{\int dy^t_{t-2\tau^{xy}}p_{N_x,\cdot}[\{t^x\}_1^{N_x},y^t_{t-2\tau^{xy}}]}.
\label{coarserate}
\end{align}
To calculate $\lambda_x$ thus requires approximation of the component integrals and probability densities. Given specific sequences of spikes in $x$ and $y$, the densities may be represented directly by eqs.~(\ref{eq:probden0}) and (\ref{probden}) with exponentiated integrals performed numerically with convergence in a discrete time parameter $\Delta t$.

The complete infinite series of integrals in Eq.~(\ref{coarserate}) of the form in Eq.~(\ref{infint}) quickly become infeasible so instead of directly computing the infinite nested integrals we choose a cutoff, $k$, for the number of spikes to include in the source path $y_{t-2\tau^{xy}}^{t}$ and then replace each of the $k$ sets of $k$ nested integrals with a separate Monte Carlo integration scheme. This is achieved, for a given $\{i\in[0,k]\}\in\mathbb{N}$, by placing $N$ spikes randomly, with floating point accuracy, on the interval $[t-2\tau^{xy},t)$ and then taking the appropriate average of the associated path probability densities. As with all Monte Carlo integration, this average does not take into account the phase space volume of the original integrals which represent the ``size'' associated with the number of ways to arrange $k$ spikes on the interval in continuous time (such that $t_y^1<t_y^2<t_y^3$ and so on). This volume is given by the integral $I_k(t-2\tau^{xy},t)$ where
\begin{align}
I_{n}(t_0,t)=\int_{t_0}^tdt_1\int_{t_1}^t dt_2\int_{t_2}^tdt_3\ldots\int_{t_{n-1}}^tdt_n,
\end{align}
which we can solve by induction, since
\begin{align}
I_{n}(t_0,t)=\int_{t_0}^tI_{n-1}(t_1,t)dt_1,
\end{align}
such that
\begin{align}
I_{n}(t_0,t)=\frac{(t-t_0)^n}{n!}.
\end{align}
We point out that one could approach the problem by constructing the limit of a discretized time space (using discretization $\delta t$, for example), thus considering probabilities, differing from the probability densities by $(\delta t)^n$ and where the phase space volume would be given by a binomial coefficient such that
\begin{align}
I_{n}(t_0,t)=\frac{(t-t_0)^n}{n!}=\lim_{\delta t\to 0}(\delta t)^n\binom{(t-t_0)/\delta t}{n}.
\end{align}
Proceeding, we may approximate the integrals
\begin{widetext}
\begin{align}
&\int dy^t_{t-2\tau^{xy}}p_{N_x,\cdot}[\{t^x\}_1^{N_x},y^t_{t-2\tau^{xy}}]\nonumber\\
&\qquad=\lim_{\substack{\Delta t\to 0\\k\to\infty\\N\to\infty}}\sum_{N_y=0}^{k}\frac{(2\tau^{xy})^{N_y}}{N_y!N}\sum_{i=1}^{N}p_{N_x,(\Delta t)}^{(0^{+},\tau^{xy})}(x^t_{t-\tau^{xy}}\equiv\{\tau^{xy},t,\{t^x\}_1^{N_x}\}|\{y^t_{t-2\tau^{xy}}\equiv\{\tau^{xy},t,[\{t^y\}_1^{N_y}]_i\}\})\nonumber\\
&\qquad\qquad\times p_{N_y,(\Delta t)}^{(0^+,0^+)}(y^t_{t-2\tau^{xy}}\equiv\{\tau^{xy},t,[\{t^y\}_1^{N_y}]_i\})
\label{bigint1}
\end{align}
and
\begin{align}
&\int dy^t_{t-2\tau^{xy}}\lambda_{x|y}^{N_x,\cdot}(y^t_{t-\tau^{xy}})p_{N_x,\cdot}[\{t^x\}_1^{N_x},y^t_{t-2\tau^{xy}}]\nonumber\\
&=\lim_{\substack{\Delta t\to 0\\k\to\infty\\N\to\infty}}\sum_{N_y=0}^{k}\frac{(2\tau^{xy})^{N_y}}{N_y!N}\sum_{i=1}^{N}\lambda_{x|y}^{N_x,N_y}(y^t_{t-2\tau^{xy}}\equiv\{\tau^{xy},t,[\{t^y\}_1^{N_y}]_i\}\})\nonumber\\
&\qquad\times p_{N_x,(\Delta t)}^{(0^{+},\tau^{xy})}(x^t_{t-\tau^{xy}}\equiv\{\tau^{xy},t,\{t^x\}_1^{N_x}\}|\{y^t_{t-2\tau^{xy}}\equiv\{\tau^{xy},t,[\{t^y\}_1^{N_y}]_i\}\}) p_{N_y,(\Delta t)}^{(0^+,0^+)}(y^t_{t-2\tau^{xy}}\equiv\{\tau^{xy},t,[\{t^y\}_1^{N_y}]_i\})
\label{bigint2}
\end{align}
\end{widetext}
where $[\{t^y\}_1^{N_y}]_i$ indicates the $i$th instance of $N_y$ randomly generated spikes in the source on the interval $[t-2\tau^{xy},t)$ and the probability densities labeled with $\Delta t$ indicate they have used $\Delta t$ as a discretization parameter in their numerical integrals. 
In our example model, where $y$ is a simple Poisson process, we have 
\begin{align}
&p_{N_y}^{(0^+,0^+)}(y^t_{t-2\tau^{xy}}\equiv\{\tau^{xy},t,[\{t^y\}_1^{N_y}]_i\})\nonumber\\
&\quad=(\lambda_y)^{N_y}\exp{\left[-\lambda_y(2\tau^{xy})\right]}\nonumber\\
&\quad=\lim_{\Delta t\to 0}p_{N_y,(\Delta t)}^{(0^+,0^+)}(y^t_{t-2\tau^{xy}}\equiv\{\tau^{xy},t,[\{t^y\}_1^{N_y}]_i\})\nonumber\\
&\quad=\lim_{\Delta t\to 0}(\lambda_y\Delta t)^{N_y}(1-\lambda_y\Delta t)^{\frac{2\tau^{xy}}{\Delta t}-N_y}.
\end{align}
The ratio of these two integrals, Eqs.~(\ref{bigint1}) and (\ref{bigint2}), then gives an estimate for $\lambda_x$ given a path history $x^t_{t-\tau^{xy}}$ containing $N_x$ spikes. We note that in practice $k$ is chosen at runtime by comparing convergence in $\lambda_x$ to a tolerance parameter while $\Delta t$ and $N$ are chosen at compile time.\\
\\
All of the above specifies how to construct $\lambda_x$ for a given path history in $x$, however, when modeling a continuous time process we wish to obtain a value at arbitrary points in time in order to meet some practical time discretization procedure. This can become infeasible and so various strategies are implemented to approximate and speed up this process. First, we assume the property in the rate functions that for any $k_x, k_y$, functions $\lambda_{x|y}(\{t^x\}_1^{k_x},\{t^y\}_1^{k_y})$ and $\lambda_{y|x}(\{t^x\}_1^{k_x},\{t^y\}_1^{k_y})$ (in the general case) are smooth in $\{t^x\}_1^{k_x},\{t^y\}_1^{k_y}$. This allows us to assume smoothness in $\lambda_x(\{t^x\}_1^k)$ for $k$ spikes in $x$ on some interval $[t-s,t)$. This combined with the observation that as time progresses $\lambda_x$ as a function of a cluster of $n$ spikes in $[t-s,t)$ is smooth in a single variable describing the relative position of the cluster in the interval until either a spike in the cluster leaves the interval or a new spike enters by virtue of $x$ spiking points towards a general interpolation scheme described below, where we focus on the special case of the utilized example where $s=\tau^{xy}$:
\begin{enumerate}
\item For phase spaces containing a manageable number, $n_x$, of spikes in $x$ on $[t-\tau_x,t)$ (e.g., 2) precompute $\lambda_x(x^t_{t-\tau^{xy}}\equiv\{\tau^{xy},t,\{t^x\}_1^{n_x}\})$ [alongside $\lambda_x(x^t_{t-\tau^{xy}}\equiv\{\tau^{xy},t\})$ being a constant value for when there are no spikes on the interval] at values $t^x_i=\{t-\tau^{xy}, t-\tau^{xy}+\Delta \tau^{\text{interp}},\ldots, t-\Delta t\}$ where $\Delta \tau^{\text{interp}}$ is a tuneable interpolation parameter. Here, $\Delta t$ is included in the final value since $t$ is the ``current'' time such that a spike at $t$ is not in the processes' history reflecting the right-open interval $[t-\tau^{xy},t)$. 
\item Numerically generate a coevolving sequence of spikes using $\lambda_{x|y}(y^t_{t-2\tau^{xy}})$ and $\lambda_{y}$ utilizing temporal discretization $\Delta t \ll \Delta \tau^{\text{interp}}$, $\Delta t \ll 1$.
\item Partition the resultant spike train in $x$ into intervals, $[t^{\text{int}}_{i-1},t^{\text{int}}_i)$, where for any $t'\in[t^{\text{int}}_{i-1},t^{\text{int}}_i)$ there are a constant number of spikes on the interval $[t'-\tau^{xy},t')$. Given the discretization scheme, there is a finite probability of a spike leaving the window to the left at the same time as a spike enters from the right after it is generated. In such cases, the regimes before and after this event are partitioned.
\item First, we consider intervals $[t^{\text{int}}_{i-1},t^{\text{int}}_i)$ where the number of spikes in that interval, $N_x$, is less than or equal to the established manageable number of spikes, $n_x$. For such values of $n_x$ we can take any such spike history and estimate $\lambda_x$ by interpolating between the precomputed $\lambda_x$ values in step one for the closest matching spike histories (based on the $\Delta \tau^{\text{interp}}$ scheme). This is performed for each required $t'\in[t^{\text{int}}_{i-1},t^{\text{int}}_i)$ according to the numerically generated spike trains with time discretization $\Delta t$.
\item Next, we consider the remaining intervals $[t^{\text{int}}_{i-1},t^{\text{int}}_i)$ such that the number of spikes in that interval, $N_x$, is greater than $n_x$. A crucial observation is that in these intervals, where the number of spikes is constant, the interspike times (i.e. $t^x_i-t^x_{i-1}$) are also constant for all times $t'\in[t^{\text{int}}_{i-1},t^{\text{int}}_i)$. This means we can parametrize the entire spike sequence by the relative position of a single spike, e.g. the time of the $N_x$th spike relative to the time in question $t'$, $t^x_{N_x}$. This can be captured by the single variable $t'-t^x_{N_x}$. Since the duration of the partitioned interval is $(t^{\text{int}}_{i}-t^{\text{int}}_{i-1})$, we can compute values of $\lambda_x$ for sequences characterized by $t'-t^x_{N_x}$ for values in $[t^{\text{int}}_i-t^{x}_{N_x}-(t^{\text{int}}_{i}-t^{\text{int}}_{i-1}),t^{\text{int}}_i-t^{x}_{N_x})$ with intervals $\Delta\tau^{\text{interp}}$. We can then use these values to interpolate values of $\lambda_x$ as measured for any time $t'\in [t^{\text{int}}_{i-1},t^{\text{int}}_i)$ which are separated by the smaller discretization parameter $\Delta t$. The appropriateness of the interpolation is assured by the initial assumptions of continuity. 
\end{enumerate}

This leaves us with intervals $[t^{\text{int}}_{i-1},t^{\text{int}}_i)$ with $N_x\leq n_x$ where we utilize a precomputed interpolation scheme of dimension up to $n_x$ and a series of independent one-dimensional interpolation schemes for each remaining $[t^{\text{int}}_{i-1},t^{\text{int}}_i)$. This allows us to estimate $\lambda_x$ for any $t'\in [t^{\text{int}}_{i-1},t^{\text{int}}_i)$ for all values of $i$ and thus for the entire spike train. We note that in practice $\Delta\tau^{\text{interp}}$ is chosen through a desired interpolation density which is rounded up when necessary to fit the variable interpolation interval lengths.
\end{appendix}

\bibliographystyle{apsrev4-1}

%
\end{document}